\documentclass[12pt]{article}
\usepackage{myjheppub}
\usepackage{graphicx,booktabs,latexsym,amssymb,amsmath,bm,epsfig,psfrag,float}
\hypersetup{pdfnewwindow=true}

\preprint{
\begin{flushright}	
	TUM-HEP-1051/16\\
	IPPP/16/66
\end{flushright}}

\title{\boldmath 
	Associated production of a top pair and a $W$ boson at next-to-next-to-leading logarithmic accuracy.\unboldmath}

\author[a]{Alessandro Broggio,}
\author[b,c]{Andrea Ferroglia,}
\author[b,c]{Giovanni Ossola,}
\author[d]{and Ben D. Pecjak}

\affiliation[a]{Physik Department T31, Technische Universit\"at M\"unchen,
James Franck-Stra{\ss}e 1, D-85748 Garching, Germany}
\emailAdd{alessandro.broggio@tum.de}

\affiliation[b]{Physics Department, New York City College of Technology,
Brooklyn, NY 11201 USA}

\affiliation[c]{The Graduate School and University Center,
The City University of New York,
New York, NY 10016  USA}
\emailAdd{aferroglia@citytech.cuny.edu}
\emailAdd{gossola@citytech.cuny.edu}

\affiliation[d]{Institute for Particle Physics Phenomenology, University of Durham,
DH1 3LE Durham, United Kingdom}
\emailAdd{ben.pecjak@durham.ac.uk}

\abstract{We consider soft gluon emission corrections to the
  production of a top-antitop pair in association with a $W$ boson
  at the Large Hadron Collider. We obtain a soft-gluon
  resummation formula for this production process which is valid up to next-to-next-to-leading logarithmic accuracy. We evaluate the soft gluon resummation formula in Mellin space by means of an in-house parton level Monte Carlo code which allows us to obtain predictions for the total cross section as well as for several differential distributions. We study the impact of the soft-gluon resummation corrections in comparison to fixed order calculations.}

\begin{document}
\newcommand{\Red}[1]{\textcolor{red}{#1}}
\newcommand{\Green}[1]{\textcolor{green}{#1}}
\newcommand{\Blue}[1]{\textcolor{blue}{#1}}
\newcommand{\Cyan}[1]{\textcolor{cyan}{#1}}
\newcommand{\Magenta}[1]{\textcolor{magenta}{#1}}
\newcommand{\alert}[1]{{\bf \Red{\boldmath[ #1 ]\unboldmath}}}

\newcommand{\be}{\begin{equation}}
\newcommand{\ee}{\end{equation}}

\newcommand{\nn}{\nonumber}
\def\ff{f\hspace{-0.3cm}f}

\def\mgamc{{{\tt \small MG5\_aMC}}}

\maketitle


\section{Introduction}
In the second half of 2015, both ATLAS and  CMS  published measurements of the total cross section for  the associated production of a top-pair
and a $W$ boson at the Large Hadron Collider (LHC) operating at a center of mass energy of $8$~ TeV. CMS measured a total $t \bar{t} W$ cross section of $382^{+117}_{-102}$~fb \cite{Khachatryan:2015sha} while ATLAS measured a total cross section of $369^{+100}_{-91}$~fb  \cite{Aad:2015eua}. The two measurements are in perfect agreement with each other but they are about $1.5~\sigma$ larger than the next-to-leading order (NLO) prediction for the cross section \cite{Garzelli:2012bn, Campbell:2012dh, Maltoni:2014zpa} in the Standard Model (SM). Furthermore, ATLAS measured a $t \bar{t} W$ cross section of $0.9 \pm 0.3$~pb at $13$~TeV \cite{ATLAS:13TeV}, in agreement with NLO SM calculations.
The associated production of a top pair and a $W$ or $Z$ boson are the two processes with the heaviest final states observed to date at the LHC. These processes are interesting because they provide direct information about the coupling of the top quark to the carriers of the weak interaction. A variety of New Physics models predict changes of these couplings with respect to their SM values. In addition, the associated production of a top pair and a $W$ boson can result in events with two leptons of the same sign, jets and missing energy. These events, which are relatively rare in the SM, are exploited in supersymmetry searches.   

Recently, some of us applied Soft Collinear Effective Theory (SCET) methods\footnote{For a didactic review, see~\cite{Becher:2014oda}.} in order to study the associated production of a top pair and a 
Higgs boson in the SM beyond NLO~\cite{Broggio:2015lya}. From the theoretical point of view, the associated production of a top pair and a $W$ boson is similar to the associated production of a top pair 
and a Higgs boson and the associated production of a top pair and a $Z$ boson.  In~\cite{Broggio:2015lya}, a resummation formula was obtained and all of the elements appearing in the formula were 
evaluated to the perturbative order needed for predictions at next-to-next-to leading logarithmic (NNLL) accuracy. Among those elements, only the hard function, which receives contributions exclusively from virtual 
corrections to the tree level partonic processes, is process dependent. The other ingredients appearing in the resummation formula, namely the soft function and the evolution matrices, are insensitive to the nature 
of the color-neutral boson produced in association to the top pair. The NLO hard functions needed in \cite{Broggio:2015lya} were obtained by modifying public codes which allow for the automated evaluation of one-loop 
virtual correction to partonic processes, namely {\tt OpenLoops}~\cite{Cascioli:2011va,Denner:2002ii, Denner:2005nn, Denner:2010tr, Denner:2014gla, Denner:2016kdg,Ossola:2007ax}, {\tt GoSam}~\cite{Cullen:2011ac, Cullen:2014yla,Binoth:2008uq,Mastrolia:2010nb, Peraro:2014cba}, and~{\tt MadLoop}~\cite{Hirschi:2011pa,Hirschi:2016mdz}.  
The information encoded in the resummation formula was employed in order to obtain results that approximate the next-to-next-to-leading order (NNLO) predictions for the $t\bar{t}H$ cross section and several differential distributions 
depending on the momenta of the final state particles. However, a numerical implementation of the resummed soft-gluon emission corrections, which is computationally more expensive than the evaluation of the 
approximate NNLO formulas, was not attempted.

In the present work, we study the resummation of soft gluon emission corrections in $t \bar{t} W$ production to NNLL accuracy. The final result of our work is a parton level Monte Carlo code which allows us to evaluate the complete resummation formula and to obtain predictions for the total cross section and several differential distributions. This program will serve as a test study for similar codes which we plan to develop in order to study NNLL resummation in $t \bar{t} H$ and   $t \bar{t} Z$ production. 
In order to build the resummation formula for the associated production and of a top pair and a $W$ boson, we need the calculation of the process dependent hard functions. We achieve this goal by suitably modifying the public codes {\tt OpenLoops} and  {\tt GoSam}.
Furthermore, we evaluate the resummation formulas in Mellin space, along the lines of what was done in the case of top-pair production in the context of an approach based on SCET methods in \cite{Pecjak:2016nee, Ferroglia:2015ivv}. The resummation of the $t \bar{t} W$ total cross section and invariant mass distribution up to NNLL accuracy was carried out in \cite{Li:2014ula} in momentum space. While the technique employed in \cite{Li:2014ula} allowed the authors to study  only the invariant mass distribution, the parton level Mone Carlo  code developed for the present work can be used to evaluate several other differential distributions.

The paper  is structured as follows. In Section~\ref{sec:soft-limit}, we introduce our notation and we review the salient features of the factorization formula.  In Section~\ref{sec:Mellinresum}, we provide details on the resummation procedure which is carried out in Mellin space.  In Section~\ref{sec:ApproxResum}, we describe the matching procedure and discuss the differences between Mellin and momentum space formulas. 
In Section~\ref{sec:NumAnalyis}, we evaluate numerically the $t \bar{t} W$ total and differential cross section at NNLL and approximate NNLO accuracy. After matching both results to the NLO cross section evaluated with {\verb|MadGraph5_aMC@NLO|}~\cite{Alwall:2014hca}, we discuss the numerical impact of the soft emission corrections beyond NLO. Finally, in Section~\ref{sec:conclusions}, we summarize the results and draw our conclusions.

\boldmath
\section{
	Soft-gluon resummation for $t\bar{t}W$ hadroproduction}
\label{sec:soft-limit}\unboldmath

The partonic process underlying the associated production of a top pair and a $W $ boson at the LHC can be written as 
\begin{align}
i(p_1) + j(p_2) \longrightarrow t (p_3) + \bar{t}(p_4) +W^{\pm}(p_5) + X \, ,
\label{eq:partonicprocess}
\end{align}
where the incoming partons $i, j \in{q, \bar{q'}}$ at lowest order; beyond tree level, also the quark gluon-channel contributes to the process. In (\ref{eq:partonicprocess}) $X$ indicates  partonic final state radiation. 
If $i$ represents a light up-type quark $u$ ($c$), then $j$ will represent a down-type light quark $d$ ($s$), 
since the $W$ boson couples to an up-type and a down-type quark. In addition 
we assume the CKM matrix to be equal to the identity matrix in such a way that there is no mixing between different quark generations.
We define the Mandelstam invariants as
\begin{gather}
\hat{s} = (p_1 + p_2)^2 = 2 p_1 \cdot p_2 \, , 
\qquad 
M^2 = (p_3 + p_4 + p_ 5)^2 \,.
\end{gather}
The soft limit (sometimes referred to as partonic threshold limit or PIM limit, from the acronym of ``Pair Invariant Mass'', where the latter  is lexically inaccurate in our case) is defined by the relation
\begin{align}
\label{eq:PIMlimit}
z \equiv \frac{M^2}{\hat{s}} \longrightarrow 1 \, ,
\end{align}
so that the unobserved final state X consists of
soft partons only. In contrast to the production threshold limit, where the
partonic center-of-mass energy approaches $2 m_t +m_W$, the partonic threshold  limit in (\ref{eq:PIMlimit}) does not
impose constraints on the velocity of massive particles in the final state. 

The factorization of the QCD cross section in the soft limit is analogous to the one described in \cite{Broggio:2015lya} for the $t \bar{t} H$ production process:
\begin{align}
\label{eq:soft-fact}
\sigma(s,m_t,m_W) =&\frac{1}{2 s} \int_{\tau_{\text{min}}}^1 \frac{d \tau}{\tau} \int_{\tau}^1 \frac{dz}{\sqrt{z}} \,
\sum_{ij} \, \ff_{ij}\left(\frac{\tau}{z},\mu\right)  \nonumber \\
& \times \int  d \text{PS}_{t \bar{t} W} \text{Tr} \left[\mathbf{H}_{ij}(\{p\},\mu)  \, \mathbf{S}_{ij}\left( \frac{M(1-z)}{\sqrt{z}}, \{p\},\mu\right)\right] \, ,
\end{align}
where 
\begin{align}
\tau = \frac{M^2}{s} \, , \qquad \tau_{\rm min} = \frac{(2m_t+m_W)^2}{s} \, ,
\end{align}
with $s$ the squared hadronic center of mass energy. In (\ref{eq:soft-fact}) and in the following,  $\{p\}$ indicates  the 
external momenta $p_1, \cdots, p_5$.
The various factors appearing in the integrand of (\ref{eq:soft-fact}) have the following meaning:
The matrix trace ${\rm Tr}[\mathbf{H}_{ij} \mathbf{S}_{ij}]$ is proportional
to the spin and color averaged squared matrix element for
$t\bar{t}W+X_s$ production through two initial-state quarks with
flavors $i$ and $j$, where $X_s$ is an unobserved final state
consisting of any number of soft gluons. The (matrix valued) hard
functions $\mathbf{H}_{ij}$ are related to color decomposed virtual
corrections to the underlying $2\to 3$ scattering process, and the
(matrix valued) soft functions $\mathbf{S}_{ij}$ are related to
color-decomposed real emission corrections in the soft limit. To
leading order in the soft limit, these soft real emission corrections
receive contributions from initial-state partons with flavor indices
$ij\in\{q\bar{q'},\bar{q}q'\}$; the ``prime'' in the superscripts indicate that the $q$ is an up-type light quark $q'$ is a down-type quark and vice versa. 
Henceforth we will refer to the channels involving quarks with the generic term ``quark annihilation" channel.  Channels involving 
initial-state partons such as $qg$ and $\bar{q}g$ are subleading in the soft limit;
in this paper we will refer  to them as the ``$qg$" channel.  The soft functions, which arise from the emission of soft gluons from the external legs of the partonic process, depend on singular
(logarithmic) plus distributions of the form %
\begin{equation}
P'_n(z) \equiv \left[\frac{1}{1-z}\ln^n \left(\frac{M^2(1-z)^2}{\mu^2 z}\right) \right]_+ \, ,
\end{equation}
as well as on the Dirac delta function $\delta(1-z)$.

The parton luminosity function $\ff$ is defined as 
 \begin{align}
 \label{eq:ff}
 \ff_{ij}\left(y, \mu\right) \equiv \int_y^1 \frac{dx_1}{x_1}   f_{i/N_1} (x_1,\mu) f_{j/N_2}\left(\frac{y}{x_1}, \mu \right) \, ,
 \end{align}
where $f_{i/N}$ is the parton distribution function (PDF) for a parton with
flavor $i$ in nucleon $N$.  

The phase space integration measure $ d \text{PS}_{t \bar{t} W}$ can be obtained from the one needed for the $t \bar{t} H$ production case, provided that one replaces the Higgs boson mass with the $W$-boson mass. This phase space was described in detail in Section 2 of \cite{Broggio:2015lya}. 
The origin of the factor $1/\sqrt{z}$ in (\ref{eq:soft-fact}) has been also discussed at length in \cite{Broggio:2015lya}; here we stress the fact that one can reabsorb a factor $\sqrt{z}$ in the definition of the soft function as follows
\begin{align}
\frac{1}{\sqrt{z}} \mathbf{S}_{ij}\left( \frac{M(1-z)}{\sqrt{z}}, \{p\},\mu\right) &= 
\frac{1}{z}\sqrt{z} \mathbf{S}_{ij}\left( \frac{M(1-z)}{\sqrt{z}}, \{p\},\mu\right) \nn \\ &\equiv
\frac{1}{z} \mathbf{S'}_{ij}\left( \frac{M(1-z)}{\sqrt{z}}, \{p\},\mu\right) \, .
\label{eq:softfuncz}
\end{align} 
Consequently, one can  rewrite (\ref{eq:soft-fact}) by factoring out of the soft function only the traditional factor $1/z$. We follow this strategy when we take the Mellin transform of (\ref{eq:softfuncz}) in order to resum soft emission corrections to NNLL accuracy in Mellin space.

In order to carry out resummation, one needs to know the soft anomalous dimension $\bm{\Gamma}_{H}$, defined through the renormalization-group equation satisfied by the hard function. These anomalous dimensions do not depend on the nature of the color-neutral boson in the final state. Consequently, the anomalous dimension needed here is the same  as the quark-annihilation channel anomalous dimension which can be found in equation (2.20) of \cite{Broggio:2015lya}.

The hard function, soft function, and soft anomalous dimension
are computed order by order by means of perturbative expansions in $\alpha_s$. In order to
carry out soft-gluon resummation to NNLL order, one needs the perturbative expansions of these elements to NLO. The quark-annihilation channel soft anomalous dimension needed here  was evaluated to NLO in \cite{Ferroglia:2009ep, Ferroglia:2009ii}. The NLO soft function for $t \bar{t} W$ production was first calculated in \cite{Li:2014ula}. The NLO hard function 
can be built starting from one-loop QCD amplitudes projected on a color basis.
The UV renormalized QCD amplitudes involve IR
divergences, which appear as poles in the limit in which the dimensional regulator $\varepsilon = (4-d)/2$ (where $d$ indicates the number of space-time dimensions)
vanishes.  One needs to subtract the residual IR poles from the color decomposed one-loop amplitudes in order to
be able to assemble the hard functions since they are finite quantities.
This is done by means of appropriate IR subtraction counterterms \cite{Ferroglia:2009ep, Ferroglia:2009ii}, following the same procedure
employed in \cite{Ahrens:2010zv} for the top-quark pair production case.

While in~\cite{Li:2014ula} the NLO hard function for $t\bar{t}W$ production was calculated  numerically by means of {\tt MadLoop},  in the present paper 
we built upon the experience gained in the calculation for the $t \bar{t} H$ NLO hard function and we evaluated the NLO hard function for $t \bar{t} W$ production by customizing two of the one-loop providers programs available on the market,  {\tt GoSam} and {\tt Openloops}.  
The numerical evaluation of the hard function for the present paper has been performed by using {\tt Openloops} 
run in combination with {\tt Collier}~\cite{Denner:2002ii, Denner:2005nn, Denner:2010tr, Denner:2014gla, Denner:2016kdg}. Results have been cross-checked  by means of  {\tt GoSam} in combination with {\tt Ninja}~\cite{Mastrolia:2012bu,vanDeurzen:2013saa, Peraro:2014cba}.   

\section{Resummation in Mellin moment space \label{sec:Mellinresum}}


By combining the information encoded in the NLO hard function and soft
function with the solution of the renormalization group (RG) equations that they satisfy, it
is possible to resum logarithms of the ratio between the hard scale
$\mu_h$ (which characterizes the hard function) and the soft scale
$\mu_s$ (which is characteristic of the soft emission) up to NNLL
accuracy. When this is done, the differential hard-scattering kernels
\begin{equation}
C_{ij}(z,\mu) \equiv  \mbox{Tr} \left[ \mathbf{H}_{ij} \left(\{p\}, \mu \right)  \mathbf{S'}_{ij}\left(\sqrt{\hat{s}}(1-z),\{p\}, \mu \right) \right] \, ,
\label{eq:Cinmomentum}
\end{equation}
(where we dropped $\{p\}$ from the list of arguments of $C_{ij}$) can be expressed in resummed form as
\begin{align}
C_{ij} \left(z,\mu_f \right) =& \exp\left[{4 a_{\gamma_\phi} (\mu_s,\mu_f)} \right]
\mbox{Tr} \biggl[ \mathbf{U}_{ij}\left(\{p\}, \mu_h,\mu_s \right) \mathbf{H}_{ij}(\{p\},\mu_h)\nonumber  \\
&\times \mathbf{U}_{ij}^\dagger \left(\{p\}, \mu_h, \mu_s \right)
\tilde{\mathbf{s}}_{ij}\left(\ln{\frac{M^2}{\mu_s}} +\partial_\eta ,\{p\},\mu_s\right) \biggr] 
\frac{e^{-2 \gamma_E \eta}}{\Gamma \left(2 \eta \right)} \frac{z ^{1/2-\eta}}{(1-z)^{1-2 \eta}} \, . \label{eq:resummedC}
\end{align}
The anomalous dimensions and evolution matrices appearing in (\ref{eq:resummedC}), as
well as the Laplace transformed soft function $\tilde{\mathbf{s}}$
are the same as in \cite{Broggio:2015lya,  Li:2014ula}.
If  the hard function and soft function are evaluated at their characteristic
scales $\mu_h$ and $\mu_s$, they are free from large logarithmic corrections and can be 
safely evaluated at a given fixed order in perturbation theory.  Large logarithmic corrections
depending on the ratio $\mu_h/\mu_s$ are resummed in the evolution matrices $\mathbf{U}$. When the resummation is carried out in momentum space, one should carefully and judiciously choose the value assigned to $\mu_h$ and especially to $\mu_s$.

While, in some instances, the logarithmic corrections depending on the ratio $\mu_h/\mu_s$ are not so large that they spoil the convergence of a fixed order expansion in $\alpha_s$, soft gluon emission effects still provide the bulk of the corrections at a given perturbative
order. In those cases, it makes sense to employ the resummed hard scattering
kernels in order to obtain approximate formulas which include all of
the terms proportional to plus distributions up to a given power of
$\alpha_s$ in fixed-order perturbation theory. This was the approach followed for example in \cite{Broggio:2015lya} for the study of $t \bar{t} H$ production.
Also in this work approximate NNLO formulas including all of the plus distributions proportional to $\alpha_s^2$ are obtained and evaluated numerically.
The results are matched to complete NLO calculations obtained by means of
{\verb|MadGraph5_aMC@NLO|} 
(indicated by \mgamc\ in the rest of the paper) \cite{Alwall:2014hca} and are labeled for convenience ``nNLO'' predictions. We are able to obtain numerical results at this level of accuracy for the total $t \bar{t} W$ cross section as well as for several differential distributions. A detailed description of what is included in the nNLO predictions, in particular in relation to the terms proportional to $\alpha_s^2 \delta(1-z)$ and to power suppressed terms which can be reconstructed in part by means of SCET based methods, can be found in Section~3 of \cite{Broggio:2015lya} and we do not repeat it here.
The matching procedure to NLO calculations is described in Section~\ref{sec:ApproxResum}.

%
However, the main result of the present paper is a numerical implementation of the resummation to NNLL accuracy. For this purpose we developed an in-house parton-level Monte Carlo code which allows us to evaluate several differential distributions in a single run. The associated production of a top pair and a $W$ boson in the soft limit involves only the quark annihilation channel; consequently the corresponding Monte Carlo code requires comparatively limited running time with respect to other processes such as $t \bar{t} H$ and $t \bar{t} Z$ production. For this reason the program we developed and optimized for this work provides a valid template which can in principle be extended to the evaluation of NNLL resummation corrections to      $t \bar{t} H$ and $t \bar{t} Z$ production. 

The NNLL resummation for the  $t \bar{t} W$ production process was already carried out in momentum space in \cite{Li:2014ula}. In that work, the authors did not develop a parton level Monte Carlo but obtained predictions for the total cross section  at the LHC for center of mass energies of $7,8,13$ and $14$~TeV and the invariant mass distribution at $8$ and $13$~TeV. In this work we decided to carry out the resummation in Mellin space, following the procedure adopted in \cite{Pecjak:2016nee, Ferroglia:2015ivv, Catani:1996yz}. In the rest of this section we describe the various elements which enter the resummation formula in Mellin space. 

The Mellin transform of a function $f$ and its inverse are defined by
\begin{align}
\widetilde{f} (N) \equiv {\mathcal M}[f](N) = \int_0^1 dx x^{N-1} f(x) \, , \quad f(x) = {\mathcal M}^{-1}[\widetilde{f}](x) = \frac{1}{2 \pi i} \int_{c-i \infty}^{c+ i \infty} d N x^{-N} \tilde{f}(N) \, .
\end{align}
The constant $c$ in the extrema of integration of the inverse Mellin transform is chosen so that the integration contour lies to the right of all singularities of the function $\tilde{f}(N)$. The total cross section of (\ref{eq:soft-fact}) 
can be rewritten as 
\begin{align}
\sigma(s, m_t, m_W) = \frac{1}{2 s} \int^1_{\tau_{\text{min}}} 
\frac{d \tau}{\tau} \frac{1}{2 \pi i} \int_{c - i \infty}^{c+ \infty} d N \tau ^{-N} \sum_{ij} \widetilde{\ff}_{ij}(N,\mu)\int  d \text{PS}_{t \bar{t} W} \widetilde{c}_{ij}(N,\mu)  \, ,
\label{eq:Mellintot}
\end{align}
where $\tilde{\ff}$ is the Mellin transform of the luminosity while 
\begin{align}
\widetilde{c}_{ij}(N,\mu) = \mbox{Tr} \left[ \mathbf{H}_{ij} \left(\{p\}, \mu \right)  \widetilde{\mathbf{s}}_{ij}\left(\ln{\frac{M^2}{\bar{N} \mu_f^2}},\{p\}, \mu \right) \right] \,. 
\label{eq:Mellinfac}
\end{align}
In (\ref{eq:Mellinfac}) we showed explicitly the parton indices $i,j$ and we neglected terms suppressed by powers of $1/N$, since the soft limit $z \to 1$ in momentum space corresponds to the limit $N \to \infty$ in moment space. Furthermore, we introduced the shorthand notation $\bar{N} = e^{\gamma_E} N$. Equation  (\ref{eq:Mellinfac}) was obtained from (\ref{eq:Cinmomentum}) by assuming that the scale $\mu$ is not chosen at the partonic level in momentum space, so that does not depend on $z$.

One can then derive and solve the RG equations for the hard and soft functions in (\ref{eq:Mellinfac}). This allows one
to evaluate the hard and soft functions at suitable values of the  hard scale $\mu_h$ and soft scale 
$\mu_s$, where
large logarithms are absent. Subsequently, one can use RG evolution to obtain the hard scattering kernels at the
factorization scale $\mu_f$. The RG-improved hard-scattering kernels can be written as
 \begin{align}
 \widetilde{c}_{ij}(N,\mu_f) =  
 \mbox{Tr} \Bigg[&\widetilde{\mathbf{U}}_{ij}(\!\bar{N},\{p\},\mu_f,\mu_h,\mu_s) \, \mathbf{H}_{ij}( \{p\},\mu_h) \, \widetilde{\mathbf{U}}_{ij}^{\dagger}(\!\bar{N},\{p\},\mu_f,\mu_h,\mu_s)
  \nn \\
 & \times \widetilde{\mathbf{s}}_{ij}\left(\ln\frac{M^2}{\bar{N}^2 \mu_s^2},\{p\},\mu_s\right)\Bigg] \,  .
 \label{eq:Mellinresum}
 \end{align}
The large logarithms of the ratio of mass scales are  exponentiated
in the evolution matrix in Mellin space, $\widetilde{\mathbf{U}}$. The evolution matrix, which enters in (\ref{eq:Mellinresum}), can be obtained in a straightforward way and it is similar to the one employed  in the study of top-pair production in the quark annihilation channel \cite{Pecjak:2016nee}: it reads
\begin{align}
\widetilde{\mathbf{U}} \left(\bar{N},\{p\},\mu_f, \mu_h, \mu_s \right) =& \exp \Biggl\{
2 S_{\Gamma_{\text{cusp}}} (\mu_h, \mu_s) - a_{\Gamma_{\text{cusp}}} (\mu_h, \mu_s) \ln 
\frac{M^2}{\mu_h^2} + a_{\Gamma_{\text{cusp}}} (\mu_f, \mu_s) \ln \bar{N}^2 
 \nonumber \\
&+ 2 a_{\gamma^\phi}(\mu_s,\mu_f) 
\Biggr\} \times \mathbf{u} \left( \{p\},\mu_h,\mu_s \right) \, .
\label{eq:Uevolution}
\end{align}
The exponential in (\ref{eq:Uevolution}) is diagonal in color space; the terms in the exponent are defined as
\begin{align}
S_{\Gamma_{\text{cusp}}} (\nu, \mu) = - \int_{\alpha_s(\nu)}^{\alpha_s(\mu)} d \alpha \frac{\Gamma_{\text{cusp}}(\alpha)}{\beta(\alpha)} \int^\alpha_{\alpha_s(\nu)} \frac{d \alpha'}{\beta(\alpha')} \, , \qquad
a_\gamma(\nu,\mu) = - \int_{\alpha_s(\nu)}^{\alpha_s(\mu)} d \alpha \frac{\gamma(\alpha)}{\beta(\alpha)} \, ,
\label{eq:an-dim}
\end{align} 
where $\gamma$ in the second equation in (\ref{eq:an-dim}) indicates either $\Gamma_{\text{cusp}}$ or $\gamma^\phi$ and the QCD $\beta$-function is defined as
\begin{align}
\beta(\alpha_s) = \frac{d \alpha_s(\mu)}{d \ln \mu} \, .
\end{align}
The color non-diagonal part of the evolution matrix in (\ref{eq:Uevolution}) is the matrix 
\begin{align}
\mathbf{u} (\{p\},\mu_h, \mu_s) = {\mathcal P} \exp{\int_{\alpha_s(\mu_h)}^{\alpha_s(\mu_s)}} \frac{d \alpha}{\beta(\alpha)} 
\bm{\gamma}^h(\alpha, \{p\}) \, .
\label{eq:ADnondiag}
\end{align}
The matrix $\bm{\gamma}^h$ is the non-diagonal part of the hard function anomalous dimension for the quark annihilation channel, which depends on the scalar products $p_i \cdot p_j$ ($i,j \in \{1, \cdots,4\}$), and consequently, on the top mass\footnote{The coefficients of the expansion of the anomalous dimensions employed in this Section in powers of $\alpha_s$ can be found for example in the appendices of \cite{Ahrens:2010zv,Becher:2014oda}.} $m_t$.

The integrals in (\ref{eq:an-dim},\ref{eq:ADnondiag}) are evaluated up to a given order in $\alpha_s$; in particular, if one is interested in carrying out the resummation to NNLL, the cusp anomalous dimension $\Gamma_{\text{cusp}}$ in (\ref{eq:an-dim}) must be evaluated to 3 loops, while $\bm{\gamma}^h$ and $\gamma^\phi$ must be evaluated to 2 loops.
When one evaluates the resummation formulas in momentum space, as it was done in \cite{Ahrens:2010zv}, the results of the integrals in (\ref{eq:an-dim},\ref{eq:ADnondiag}) are expressed in terms of $\alpha_s(\mu_f)$, $\alpha_s(\mu_s)$, and $\alpha_s(\mu_h)$. This is convenient since in that approach the soft scale is a real number depending on the event kinematics. When the resummation is carried out in Mellin space instead, the soft scale is chosen to be $\mu_s \sim M/\bar{N}$, and it is a complex number since one will ultimately need to integrate over the complex $N$ variable. For this reason, it is more convenient to re-express  $\alpha_s(\mu_f)$ and $\alpha_s(\mu_s)$ in terms of $\alpha_s(\mu_h)$ by using the perturbative evolution of the strong coupling constant up to 3-loop order, which can be found for example in \cite{Moch:2005ba}. In this way, the large logarithms of the ratio $\mu_h/\mu_s$ appear explicitly in the formula for the evolution matrix, which reads 
%
\begin{align}
\widetilde{\mathbf{U}}\left(\bar{N}, \{p\},\mu_f, \mu_h, \mu_s \right) =& \exp \Biggl\{
\frac{4 \pi}{\alpha_s(\mu_h)} g_1 \left( \lambda, \lambda_f \right) +  g_2 \left( \lambda, \lambda_f \right) + \frac{\alpha_s(\mu_h)}{4 \pi}  g_3 \left( \lambda, \lambda_f \right) + \cdots \Biggr\} \nn \\
& \times \mathbf{u} ( \{p\},\mu_h, \mu_s)\, ,
\end{align} 
where 
\begin{align}
\lambda = \frac{\alpha_s(\mu_h)}{2 \pi} \beta_0 \ln{\frac{\mu_h}{\mu_s}} \, ,
\qquad
\lambda_f = \frac{\alpha_s(\mu_h)}{2 \pi} \beta_0 \ln{\frac{\mu_h}{\mu_f}}\, .
\end{align}
The explicit expressions of the leading logarithmic (LL) function $g_1$, the next-to-leading logarithmic (NLL) function   $g_2$, and the NNLL function $g_3$ can be easily derived.

The resummed formula for the hard scattering kernel to all orders, (\ref{eq:Mellinresum}), is independent of the 
hard and soft scales. However a residual dependence on those scales is present in any numerical evaluation of (\ref{eq:Mellinresum}), which must rely upon a truncation of the various factors at a given logarithmic accuracy. 
In order to keep the hard and soft functions free from large logarithms, it is reasonable to chose $\mu_h \sim M$ and, as mentioned above,  $\mu_s \sim M/\bar{N}$. With this choice of $\mu_s$ however, one runs in a well known problem: a branch cut for large values of $N$ is found in the resummed hard scattering kernel. This branch cut is related to the existence of a Landau pole in the running of $\alpha_s$. The branch cut leads to ambiguities in the choice of the integration path when one evaluates the inverse Mellin transform; in this work we choose the integration path following the \emph{Minimal Prescription} (MP) of \cite{Catani:1996yz}.

\section{Approximate and resummed formulas} \label{sec:ApproxResum}

In this Section we describe in some detail the difference between the various implementations of the approximate and resummed formulas which we evaluate numerically in Section~\ref{sec:NumAnalyis}. 

Approximate formulas for the hard scattering kernel in momentum space can be easily obtained from (\ref{eq:resummedC}) by first setting $\mu_h = \mu_s = \mu_f$. In this way, the evolution matrices $\mathbf{U}$ become equal to the identity matrix, and the factor $\eta$ vanishes. Therefore in this case the hard scattering kernel can be written as
\begin{align}
C_{ij} \left(z,\mu_f \right) =&
\mbox{Tr} \Biggl[ \mathbf{H}_{ij}(\{p\},\mu_f) 
\tilde{\mathbf{s}}_{ij}\left(\ln{\frac{M^2}{\mu_f}} +\partial_\eta ,\{p\},\mu_f\right) \Biggr] \left.
\frac{e^{-2 \gamma_E \eta}}{\Gamma \left(2 \eta \right)} \frac{z ^{1/2-\eta}}{(1-z)^{1-2 \eta}} \right|_{\eta \to 0} \, . \label{eq:appnNLOC}
\end{align}
By evaluating the hard and soft function to NLO it is possible to obtain approximate NLO formulas, which include the complete set of plus distribution of the form 
\begin{align}
P_n (z) \equiv \left[ \frac{\ln^n(1-z)}{1-z} \right]_+  \, ,
\label{eq:distP}
\end{align}
(with $n = 0,1$) and the delta function of argument $(1-z)$ at ${\mathcal O}(\alpha_s)$ with respect to the tree level. By employing the RG equation satisfied by the soft function and (\ref{eq:appnNLOC}), one can also obtain approximate NNLO formulas of the same kind as the one obtained for $t \bar{t} H$ production in \cite{Broggio:2015lya}. At ${\mathcal O}(\alpha_s^2)$ with respect to the tree level, the approximate NNLO formulas include the complete coefficients of the distribution functions in  (\ref{eq:distP}) (for $n=0, \cdots, 3$), as well as the scale dependent part of the coefficient of the delta function. The approximate formulas include also a subset of the terms which are subleading in the soft limit. Since these subleading terms are exactly of the same kind as the ones included in the formulas for  $t \bar{t} H$ production found in \cite{Broggio:2015lya},  we refer the interested reader to Section 3 in that paper for a detailed description of the origin of these subleading terms. Finally, the approximate NNLO formulas are matched to the complete NLO calculations carried out with {\mgamc}. For example, for the total cross section the matched prediction is obtained as follows:
\begin{align}
\sigma^{\text{nNLO}} = \sigma^{\text{NLO}} + \sigma^{\text{approx. NNLO}} - \sigma^{\text{approx. NLO}} \, .
\label{eq:NNLOmatching}
\end{align}
The last term in (\ref{eq:NNLOmatching}) avoids double counting of NLO terms proportional
to plus distributions and delta functions, which are included in both the NLO and the approx. NNLO cross section. All of the terms in (\ref{eq:NNLOmatching}) must be evaluated with NNLO PDFs.  In (\ref{eq:NNLOmatching}) and in the following we indicate matched NLO + approx. NNLO calculations
with the symbol “nNLO”.

Similar approximate NNLO formulas can be obtained from the hard scattering kernel in Mellin space, which can be found in  (\ref{eq:Mellinresum}). Also in this approach, in order to obtain an approximate NNLO formula one needs  to exploit the RG equation satisfied by the soft function to determine the prefactors of the logarithms of the soft scale to NNLO, and then one must set $\mu_h = \mu_s = \mu_f$ in   (\ref{eq:Mellinresum}). Ultimately, it is necessary to calculate the inverse Mellin transform of the powers of $\ln \bar{N}$ found in the approximate formula in Mellin space.
In particular, the inverse Mellin transform from Mellin space to $z$ space leads to the following replacements
\begin{align}
1 \rightarrow &\,  \delta \left(1-z\right) \, , \nn \\
\ln (\bar{N}^2) \rightarrow &\,   -2 P^0_{\ln}(z) + 2 \gamma_E\delta \left(1-z\right)  \, , \nn \\
\ln^2 (\bar{N}^2) \rightarrow &\,   4 P^1_{\ln}(z) + 4 \gamma^2_E\delta \left(1-z\right)\, , \nn \\
\ln^3 (\bar{N}^2) \rightarrow &\,   -6 P^2_{\ln}(z) + 4 \pi^2 P^0_{\ln}(z) +8 \gamma^3_E \delta(1-z)\, , \nn \\
\ln^4 (\bar{N}^2) \rightarrow &\,   8 P^3_{\ln}(z) - 16 \pi^2 P^1_{\ln}(z) + 128\zeta(3)P^0_{\ln}(z) + 16 \gamma^4_E \delta(1-z)\, ,
\end{align}
where
\begin{align}
P^n_{\ln}(z) \equiv \, \left[ \frac{\ln^n\left(\ln^2(z)\right)}{-\ln(z)}	\right]_+ .
\end{align}

The approximate NNLO formulas obtained from the Mellin space approach can then be matched to the complete NLO calculations through (\ref{eq:NNLOmatching}). In Section \ref{sec:NumAnalyis} we show numerically that the evaluation  of the nNLO formulas in the two approaches illustrated above leads to results which are numerically very close to each other; this is due to the fact that \cite{Becher:2007ty}
\begin{align}
\frac{1}{2 \pi i} \int_{c - i \infty}^{c+ i \infty} \!\!\! dz\,  z^{-N} \bar{N}^{- 2 \eta} =
\frac{e^{-2 \gamma_E \eta}}{\Gamma \left(2 \eta \right)} 
\frac{z^{1/2 - \eta}}{(1-z)^{1-2 \eta}} \left[1+ {\mathcal O} \left( (1-z)^2\right) \right] \, . \label{eq:Mellinantisoft}
\end{align}
One can recognize the $z$ dependent factor at the end of (\ref{eq:appnNLOC}) in the r.h.s. of (\ref{eq:Mellinantisoft}).

The main goal of our work is to evaluate numerically the total and differential $t \bar{t} W$ cross section to NLO+NNLL. Consequently we developed an in-house parton level Monte Carlo code which can evaluate numerically the hard scattering kernel (\ref{eq:Mellinresum}) and subsequently evaluates the inverse Mellin transform of 
(\ref{eq:Mellintot}). The number obtained in this way is the NNLL total cross section, which depends on the specific choice made for the scales $\mu_f, \mu_s$ and $\mu_h$.
By introducing the quantity $\kappa_i = \mu_i/\mu_{i,0}$, where $i = f,s,h$ and $\mu_{i,0}$ is the default choice for the scale $\mu_i$, we can indicate the NNLL cross section by
\begin{displaymath}
\sigma^{{\text{NNLL}}} \left(\kappa_f,\kappa_s,\kappa_h \right) \, .
\end{displaymath}
In order to match the NNLL resummed cross section to the NLO cross section one needs the approximate NLO cross section in Mellin space, which can be evaluated as discussed at the beginning of this section, as well as the full NLO cross section evaluated with 
\mgamc. The latter two quantities will depend on $\kappa_f$. Finally, NLO+NNLL predictions are obtained as follows
\begin{align}
\sigma^{{\text{NLO+NNLL}}} \left(\kappa_f,\kappa_s,\kappa_h \right) =& 
\sigma^{{\text{NLO}}} \left(\kappa_f\right) + 
\sigma^{{\text{NNLL}}} \left(\kappa_f,\kappa_s,\kappa_h \right) - 
\sigma^{{\text{approx. NLO}}} \left(\kappa_f\right)  \, .
\label{eq:NLOpNNLLmatching}
\end{align}
By subtracting the last term in (\ref{eq:NLOpNNLLmatching}) one avoids the double counting of the tree level and of the terms proportional to the plus distributions and delta function at NLO, which are included in both the NLO and the NNLL cross section. As in (\ref{eq:NNLOmatching}), all of the terms in (\ref{eq:NLOpNNLLmatching}) must be evaluated with NNLO PDFs.

Finally, in order to asses the impact of the resummed terms which are of ${\mathcal O} (\alpha_s^3)$ and higher with respect to the tree level, we also evaluate the NNLL formulas expanded to NNLO. These calculations, which appear only in Table~\ref{tab:CSWp8} and Figure~\ref{fig:Wp8NNLOrats}, are carried out simply by re-expanding the various elements in (\ref{eq:Mellinresum}) to NNLO, and by then taking the inverse Mellin transform. These results are different from the ones obtained from the nNLO formulas described above, since they do depend on the choice of $\mu_s$ and $\mu_h$ as well as $\mu_f$.

\section{Numerical analysis \label{sec:NumAnalyis}}
\label{sec:pheno}

With the formalism described in the previous sections, we are able to calculate the cross section for the associated production of a top pair and a $W$ boson both to approximate NNLO and to NNLL accuracy. Numerical predictions for the total cross section and several differential distributions were obtained by means of a dedicated {\tt Fortran} program. 
With respect to the code which was developed for the study of the associated production of a top pair and a Higgs boson~\cite{Broggio:2015lya}, the present program has been enhanced in order to evaluate the complete NNLL resummation formulas and not only the nNLO corrections. 

In the following two subsections we collect our results for the total cross section and for some inclusive differential distributions, respectively. In both cases, we first try to assess to what extent the soft emission limit $M \to \sqrt{\hat{s}}$ provides a good approximation of the known complete NLO results. Experimentally measurable observables such as the total cross section
or differential distributions at their peaks are also sensitive to regions of phase space
far away from the soft limit. However,  the soft limit is dominant
also in those cases if the mechanism of dynamical threshold enhancement \cite{Becher:2007ty,Ahrens:2010zv}
occurs. This simply means that the parton luminosities appearing in (\ref{eq:soft-fact}) must drop off quickly enough away from the soft limit region,  so that an expansion
under the integral of the partonic cross section in the soft limit is justified.
For this reason we compare approximate NLO results with complete NLO results obtained from \mgamc. The approximate NLO results are obtained by re-expanding the NNLL
resummed partonic cross section to NLO; consequently they reproduce completely all
of the terms singular in the $z \to 1$ limit in the NLO partonic cross section, but they miss terms which are subleading in the soft limit. In all cases considered in this work, we 
observe that the soft approximation
works quite well at NLO. While this fact does  not immediately imply that the
same is true at NNLO, it is an important sanity check nonetheless. 
In addition, we also study the comparison of the approximate NLO predictions with the NLO calculations carried out by excluding the contribution of the quark-gluon channel. In fact, the quark-gluon channel, which contributes to the observables starting from NLO,  is subleading in the soft limit and as such it cannot be reproduced by the approximate formulas. We found that the contribution of the quark-gluon channel to the NLO cross section is small but not completely negligible numerically. In particular, the quark gluon channel changes 
slightly the shape of the distributions. As expected, the NLO cross section without the quark gluon channel contribution is in very good agreement with the approximate NLO calculations in the soft limit.

\begin{table}[t]
	\begin{center}
		\def\arraystretch{1.3}
		\begin{tabular}{|c|c||c|c|}
			\hline $M_W$ & $80.385$ & $m_t$ & $173.2$~GeV\\ 
			\hline $M_Z$ &  $91.1876$~GeV & $m_H$ & $125$~GeV \\ 
			\hline $1/\alpha$ & $137.036$ & $\alpha_s \left(M_Z\right)$ & from MMHT 2014 PDFs \\ 
			\hline 
		\end{tabular} 
		\caption{Input parameters employed throughout the calculation. \label{tab:tabGmu}}
	\end{center}
\end{table}

After these preliminary studies, we present the cross section and differential distributions at nNLO and NNLL accuracy matched to the full NLO prediction, which represent the main results of this work.
It is important to observe that the contribution of the quark-gluon channel at NLO is included in both the nNLO and the NLO+NNLL predictions through the matching procedure.
Since the associated production of top-quark pair and a $W$ boson was already measured at the LHC operating at both a center of mass energy of $8$~TeV as well as at $13$~TeV
\cite{Khachatryan:2015sha, Aad:2015eua, ATLAS:13TeV}, we consider both collider energies in our analysis.  The parameters we employed in the numerical calculations are summarized in Table~\ref{tab:tabGmu}.

We discuss now the choice of the scales employed in the numerical calculations. Since all of our calculations are based upon a factorization formula derived in PIM kinematics, we set the factorization and renormalization scale in all  calculations equal to 
the invariant mass $M$ of the massive particles in the final state (i.e. the top pair and the $W$ boson). We follow the traditional procedure of estimating the perturbative uncertainty associated to the missing higher order corrections by varying the renormalization/factorization scale in the range $[M/2, 2M]$. 

In the resummed formulas, however, one also needs to choose the hard scale $\mu_h$ and the soft scale $\mu_s$. Following the approach employed in \cite{Ferroglia:2015ivv,Pecjak:2016nee} for the case of top-quark pair production, we choose $\mu_h = M$ and $\mu_s = M/\bar{N}$. 
With this choice of $\mu_s$,  SCET resummation is performed
at the level of Mellin-space partonic cross sections; the scales resummed here are the same ones which enter in the ``direct QCD'' approach to resummation \cite{Bonvini:2012az, Bonvini:2014qga}.
The scale uncertainty associated with the NLO+NNLL predictions is evaluated as follows: Every scale in the resummed partonic cross section is separately varied in the interval $[\mu_{i,0}/2, 2 \mu_{i,0}]$, where $i \in \{f,h,s\}$ and the subscript ``$0$'' indicates the reference value chosen for that scale.
Subsequently, for each observable (the total cross section or the value of the cross section in a particular bin of a differential distribution) one evaluates 
\begin{align}
\Delta O^+_i &= \text{max}\{O(\kappa_i = 1/2,\kappa_i = 1,\kappa_i =2) \} - \bar{O} \, , \nn \\
\Delta O^-_i &= \text{min}\{O(\kappa_i = 1/2,\kappa_i = 1,\kappa_i =2) \} - \bar{O} \, ,
\end{align}
where $\kappa_i = \mu_i/\mu_{i,0}$, while $\bar{O}$ indicates  the observable evaluated by setting all of the scales to their default value ($\kappa_i = 1$ for $i \in \{f,h,s\}$). Finally, $\Delta O^+_f,\Delta O^+_s,  \Delta O^+_h$ are combined in quadrature in order to obtain the scale uncertainty above the central value $\bar{O}$. In the same way, $\Delta O^-_f,\Delta O^-_s,  \Delta O^-_h$ are combined in quadrature in order to obtain the scale uncertainty below the central value $\bar{O}$.

\subsection{Total cross section}
\label{sec:xs}
%

\begin{table}[t]
	\begin{center}
		\def\arraystretch{1.3}
		\begin{tabular}{|c|c|c|c|}
			\hline  order & PDFs order & code & $\sigma$ [fb]\\ 
			\hline LO & LO & \mgamc & $82.0^{+21.2}_{-15.7}$ \\
			\hline NLO & NLO & \mgamc & $121.6^{+15.2}_{-14.0}$ \\
			\hline NLO no $qg$ & NLO & \mgamc & $118.1^{+10.3}_{-11.3}$ \\
			\hline app. NLO & NLO & in-house MC & $116.0^{+10.3}_{-11.6}$ \\
			\hline nNLO (momentum) & NNLO & in-house MC +\mgamc  & $127.7^{+9.2}_{-7.4}$ \\
			\hline nNLO (Mellin) & NNLO & in-house MC +\mgamc  & $127.6^{+9.2}_{-7.4}$ \\
			\hline NLO+NLL & NLO&  in-house MC  +\mgamc  & $124.8^{+13.1}_{-8.0}$ \\
			\hline (NLO+NNLL)$_{\rm NNLO \, exp.}$  & NNLO & in-house MC +\mgamc  & $126.7^{+5.0}_{-6.5}$ \\
			\hline NLO+NNLL  & NNLO & in-house MC +\mgamc  & $128.7^{+5.5}_{-4.7}$ \\
			\hline 
		\end{tabular} 
		\caption{Total cross section  for $t \bar{t} W^+$ at the LHC with $\sqrt{s} = 8$~TeV and MMHT 2014 PDFs \cite{Harland-Lang:2014zoa}. The uncertainties reflect scale variations. 
\label{tab:CSWp8}}	
	\end{center}
\end{table}
 The values of the total cross section for the production of a $t \bar{t}$ pair and a $W^+$ boson at the LHC operating at a center of mass energy of $8$~TeV are shown in Table~\ref{tab:CSWp8}. By comparing the 3rd and 5th line of that table one can observe that the complete NLO cross section is larger than the approximate NLO cross section by $ 4.8 \%$. The NLO cross section without the quark-gluon channel contribution, which can be found in the 4th line of the table, is slightly smaller than the complete NLO cross section. The approximate NLO cross section (5th line) and the NLO cross section without the quark-gluon channel contribution (4th line) differ by only $1.8 \%$. The scale uncertainty found in the approximate NLO calculation is similar to the uncertainty affecting the full NLO calculation, but also almost identical to the scale uncertainty found by excluding the contribution of the quark-gluon channel from the NLO prediction. Indeed, it is a known fact that the NLO quark-gluon channel contribution has a relatively large impact on the NLO scale uncertainty. We stress once more the fact that contribution of the quark-gluon channel at NLO is included in both the nNLO and the NLO+NNLL predictions through the matching procedure.   

\begin{table}[t]
	\begin{center}
		\def\arraystretch{1.3}
		\begin{tabular}{|c|c|c|c|}
			\hline  order & PDFs order & code & $\sigma$ [fb]\\ 
			\hline LO & LO & \mgamc & $ 37.3^{+9.7}_{-7.2}$ \\
			\hline NLO & NLO & \mgamc & $53.7^{+6.8}_{-6.3}$ \\
			\hline NLO no $qg$ & NLO & \mgamc & $52.3^{+4.5}_{-5.0}$ \\
			\hline app. NLO & NLO & in-house MC & $ 51.3^{+4.8}_{-5.0}$ \\
			\hline nNLO (momentum) & NNLO & in-house MC +\mgamc  & $ 57.7^{+4.1}_{-3.3}$ \\
			\hline nNLO (Mellin) & NNLO & in-house MC +\mgamc  & $57.7^{+4.1}_{-3.3}$ \\
			\hline NLO+NNLL  & NNLO & in-house MC +\mgamc  & $ 58.4^{+2.4}_{-2.2}$ \\
			\hline 
		\end{tabular} 
		\caption{Total cross section  for $t \bar{t} W^-$ at the LHC with $\sqrt{s} = 8$~TeV and MMHT 2014 PDFs.   \label{tab:CSWm8}}	
	\end{center}
\end{table}

\begin{table}
	\begin{center}
		\def\arraystretch{1.3}
		\begin{tabular}{|c|c|c|c|}
			\hline  order & PDFs order & code & $\sigma$ [fb]\\ 
			\hline LO & LO & \mgamc & $ 202.1^{+45.5}_{-34.9}$ \\
			\hline NLO & NLO & \mgamc & $316.9^{+39.3}_{-34.9}$ \\
			\hline NLO no $qg$ & NLO & \mgamc & $293.3^{+19.3}_{-22.7}$ \\
			\hline app. NLO & NLO & in-house MC & $ 288.1^{+21.4}_{-23.8}$ \\
			\hline nNLO (Mellin) & NNLO & in-house MC +\mgamc  & $ 330.5^{+26.2}_{-19.2}$ \\
			\hline NLO+NNLL  & NNLO & in-house MC +\mgamc  & $ 333.0^{+14.9}_{-12.4}$ \\
			\hline 
		\end{tabular} 
		\caption{Total cross section  for $t \bar{t} W^+$ at the LHC with $\sqrt{s} = 13$~TeV and MMHT 2014 PDFs.  \label{tab:CSWp13}}	
	\end{center}
\end{table}

\begin{table}
	\begin{center}
		\def\arraystretch{1.3}
		\begin{tabular}{|c|c|c|c|}
			\hline  order & PDFs order & code & $\sigma$ [fb]\\ 
			\hline LO & LO & \mgamc & $ 105.4^{+23.5}_{-18.2}$ \\
			\hline NLO & NLO & \mgamc & $161.9^{+20.4}_{-18.1}$ \\
			\hline NLO no $qg$ & NLO & \mgamc & $149.3^{+9.2}_{-11.2}$ \\
			\hline app. NLO & NLO & in-house MC & $ 147.6^{+10.5}_{-11.9}$ \\
			\hline nNLO (Mellin) & NNLO & in-house MC +\mgamc  & $ 171.8^{+13.3}_{-9.7}$ \\
			\hline NLO+NNLL  & NNLO & in-house MC +\mgamc  & $173.1^{+7.7}_{-6.0}$ \\
			\hline 
		\end{tabular} 
		\caption{Total cross section  for $t \bar{t} W^-$ at the LHC with $\sqrt{s} = 13$~TeV and MMHT 2014 PDFs.   \label{tab:CSWm13}}	
	\end{center}
\end{table}
Very similar conclusions can be drawn by comparing complete NLO results, NLO results without the quark-gluon channel, and approximate NLO results for the production of a top pair and a $W^-$ boson at $8$~TeV (Table~\ref{tab:CSWm8}), as well as for the production of $t \bar{t} W^+$ at $13$~TeV (Table~\ref{tab:CSWp13}), and for the   production of 
$t \bar{t} W^-$ at $13$~TeV (Table~\ref{tab:CSWm13}).  These observations motivate us to carry out the analysis of the total cross section to nNLO and NLO+NNLL.

The nNLO calculations can be either carried out in momentum space (where the soft function is calculated in Laplace space), following the same procedure employed for the associated top-pair and Higgs boson production in \cite{Broggio:2015lya}, or by re-expanding the NNLL resummation formulas in Mellin space to NNLO after setting all scales equal and by subsequently performing an inverse Mellin transform. For the LHC operating at a center of mass energy of $8$~TeV we followed both procedures in order to compare the results. As it can be seen by looking at Tables~\ref{tab:CSWp8} and \ref{tab:CSWm8}, the nNLO total cross section calculated in momentum space (6th line) and Mellin space (7th line) are almost identical for what concerns both the central value and the scale uncertainty. This fact is not unexpected, because it is known that the soft function, as written in (\ref{eq:soft-fact}) and (\ref{eq:softfuncz}), and the soft function obtained by calculating the inverse Mellin transform of the partonic cross section in Mellin space, differ only by terms of order $(1-z)^2$. For this reason in the following we consider only nNLO calculations carried out starting from the resummation formula in Mellin space.
For all of the cases which we consider in this work ($t\bar{t}W^+$ and $t\bar{t}W^-$ production at both $\sqrt{s} = 8$~TeV and $\sqrt{s}=13$~TeV) the nNLO cross section is $4$ to $7~\%$ larger than the NLO one, and it is affected by a residual scale uncertainty which is roughly a bit more than half of the one affecting the NLO cross section.

The total cross section at NLO+NNLL can be found in the last line of Tables~\ref{tab:CSWp8}-\ref{tab:CSWm13}. The NLO+NNLL cross section is always slightly larger than the nNLO one, and it is $5$ to $9 \%$ larger than the corresponding NLO cross section. The residual scale uncertainty affecting NLO+NNLL calculations is in all cases a bit smaller than the one found at nNLO. At the same time the NLO+NNLL cross sections has a residual scale uncertainty which is $\sim 35 \%$ of the NLO scale uncertainty. For $t \bar{t} W^+$ production at $8$~TeV (Table~\ref{tab:CSWp8}) we also compared the NLO+NNLL and the NLO+NLL calculations; we observe that the NLO+NNLL cross section is larger than the NLO+NLL cross section by about $3 \%$, while, as expected, the perturbative uncertainty at NLO+NNLL is slightly smaller than the perturbative uncertainty at  NLO+NLL.
By comparing the last two lines of Table~\ref{tab:CSWp8} it is possible to assess the impact of the resummed corrections beyond NNLO. The NLO+NNLL cross section is larger than the NLO+NNLL expanded to NNLO by less than 2\%. The scale uncertainty interval in the resummed prediction is marginally smaller than the expanded one.

In summary, we can conclude that the soft emission corrections accounted for in the nNLO and NLO+NNLL calculations of the total cross section for the associated production of a top pair and a $W$ boson lead to a moderate increase the central value of the cross section, which remains in the scale uncertainty bracket obtained from NLO calculations. At the same time a moderate decrease of the residual scale uncertainty is observed when comparing nNLO and NLO+NNLL predictions with the corresponding cross section evaluated at  NLO.

\subsection{Differential distributions}
\label{sec:dxs}

In principle, the approach adopted 
in this work allows us to calculate any differential distribution which depends on the momenta of the massive particles in the final state. We evaluate some of these distributions by employing standard Monte Carlo methods. In
particular, when we evaluate the nNLO or NLO+NNLL corrections to the total
cross section in (\ref{eq:soft-fact}), we use the phase-space and four-momenta parameterizations
described in Section 2  of \cite{Broggio:2015lya} in order to obtain predictions for  binned distributions.
In this subsection we consider the following differential distributions:

\begin{itemize}
	\item  Distribution differential with respect to the invariant mass of the massive final state particles, $M$.
	\item Distribution differential with respect to the invariant mass of the top-quark pair, $M_{t\bar{t}}$.
	\item Distribution differential with respect to the transverse momentum of the $W$ boson, $p_T^W$.
	\item Distribution differential with respect to the transverse momentum of the top quark, $p_T^t$.
\end{itemize}
The scale choices employed in the evaluation of the various distributions presented below have been described at the beginning of Section~\ref{sec:NumAnalyis}. While these choices are particularly suitable for the evaluation of the final state invariant mass distribution, for simplicity we evaluate all of the distributions that we discuss here with the same scale choices.
Since detailed phenomenological analyses might require different scale choices for differential distributions other than the invariant mass distribution, our code can evaluate observables for an arbitrary choice (fixed or dynamical) of the scales. However, since we get all of the differential distributions in a single run of the parton-level Monte Carlo, 
we decided to evaluate them all by employing the same scale choices. 
In addition, to evaluate each differential distribution with different scale choices would require a much  longer running time.

\begin{figure}[tp]
	\begin{center}
		\begin{tabular}{cc}
			\includegraphics[width=7.2cm]{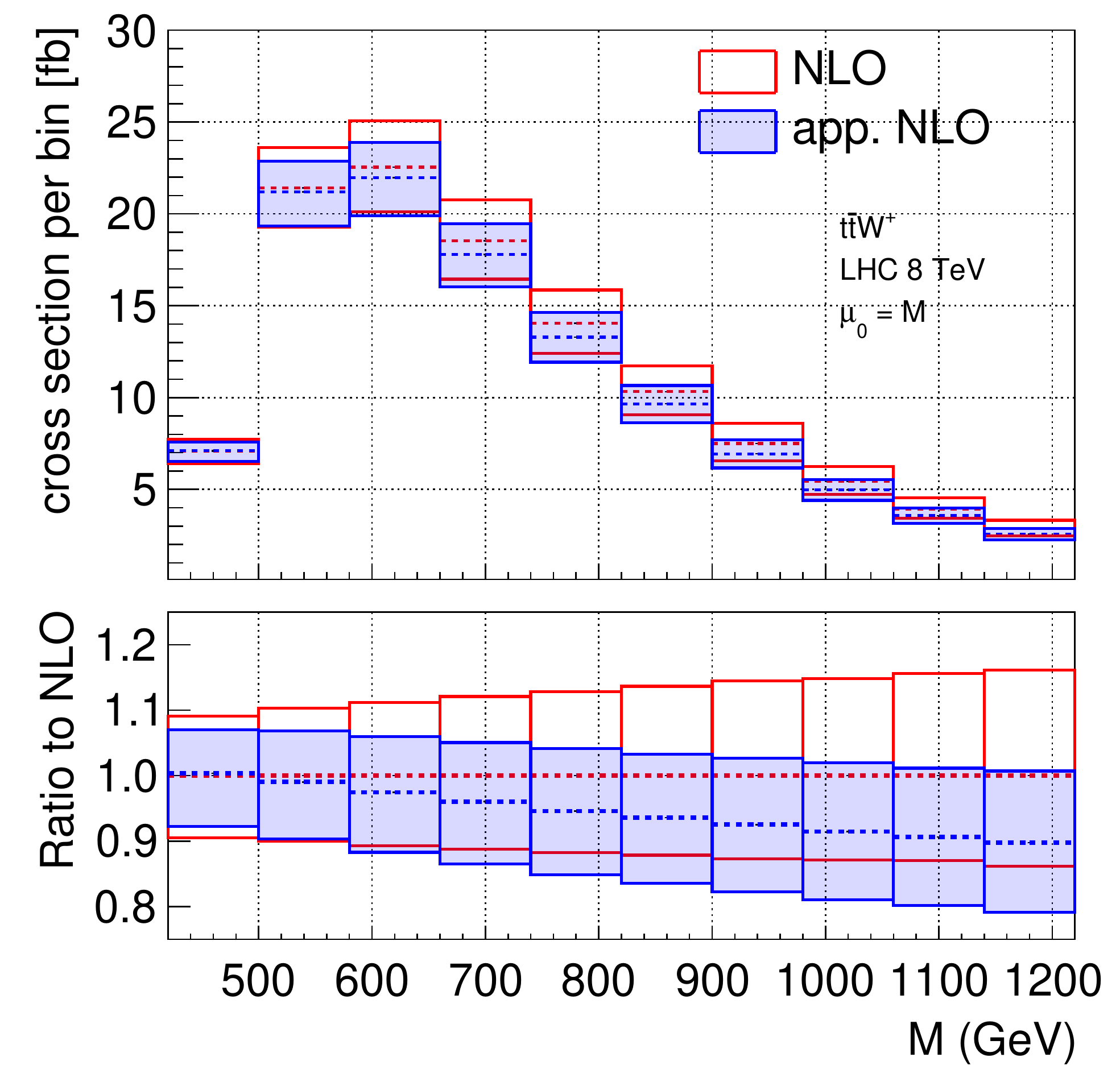} & \includegraphics[width=7.2cm]{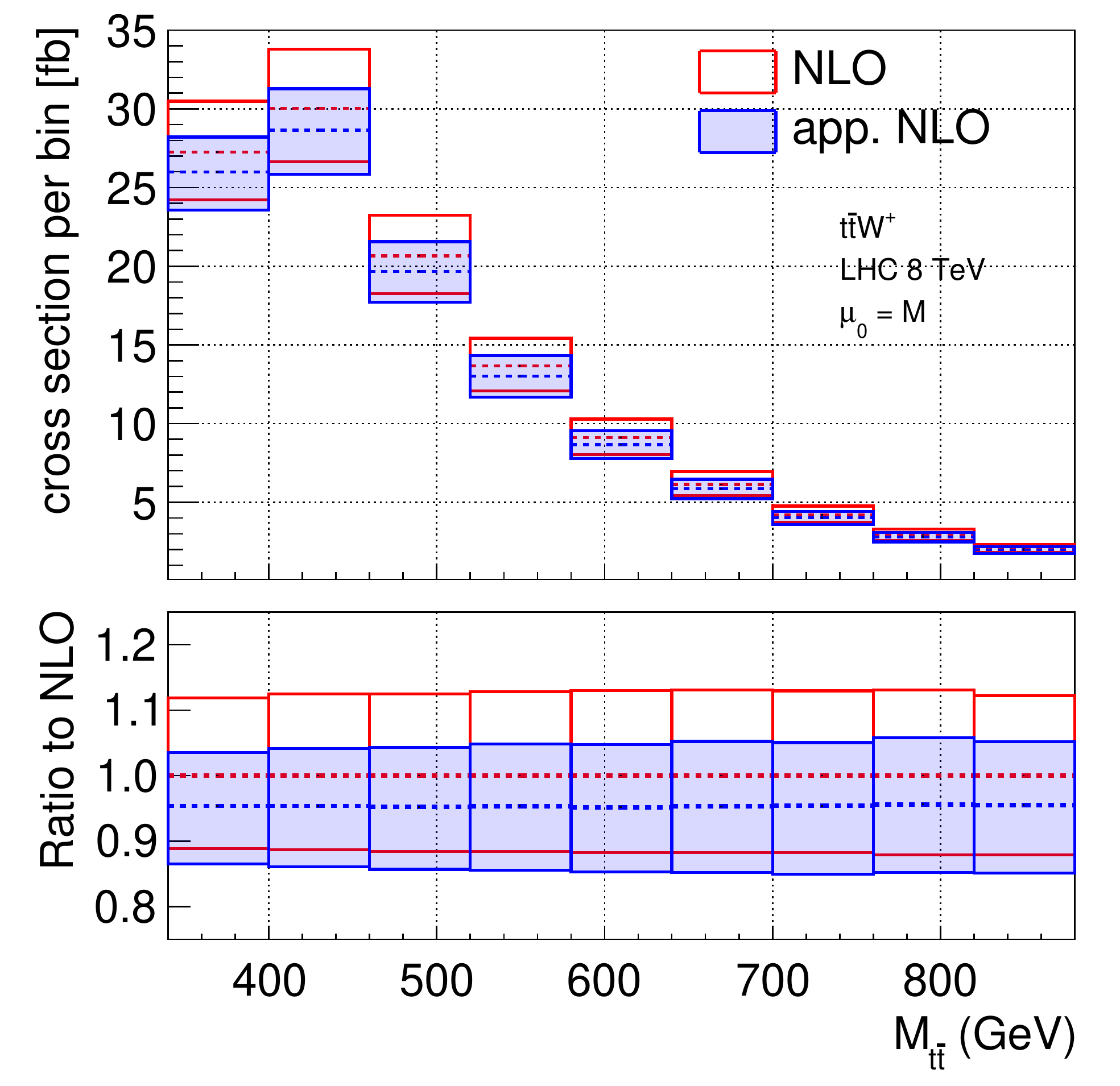} \\
			\includegraphics[width=7.2cm]{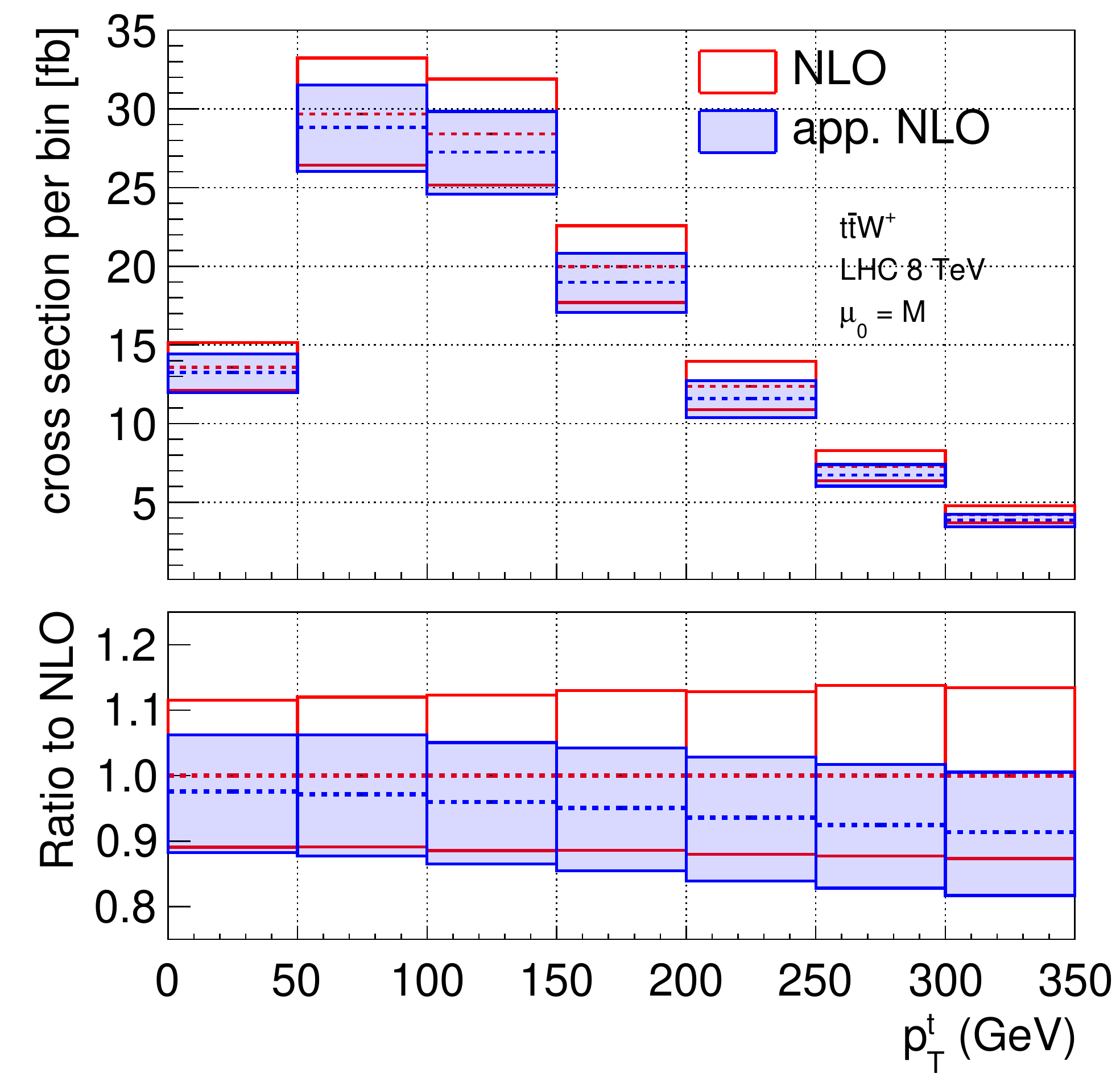} & \includegraphics[width=7.2cm]{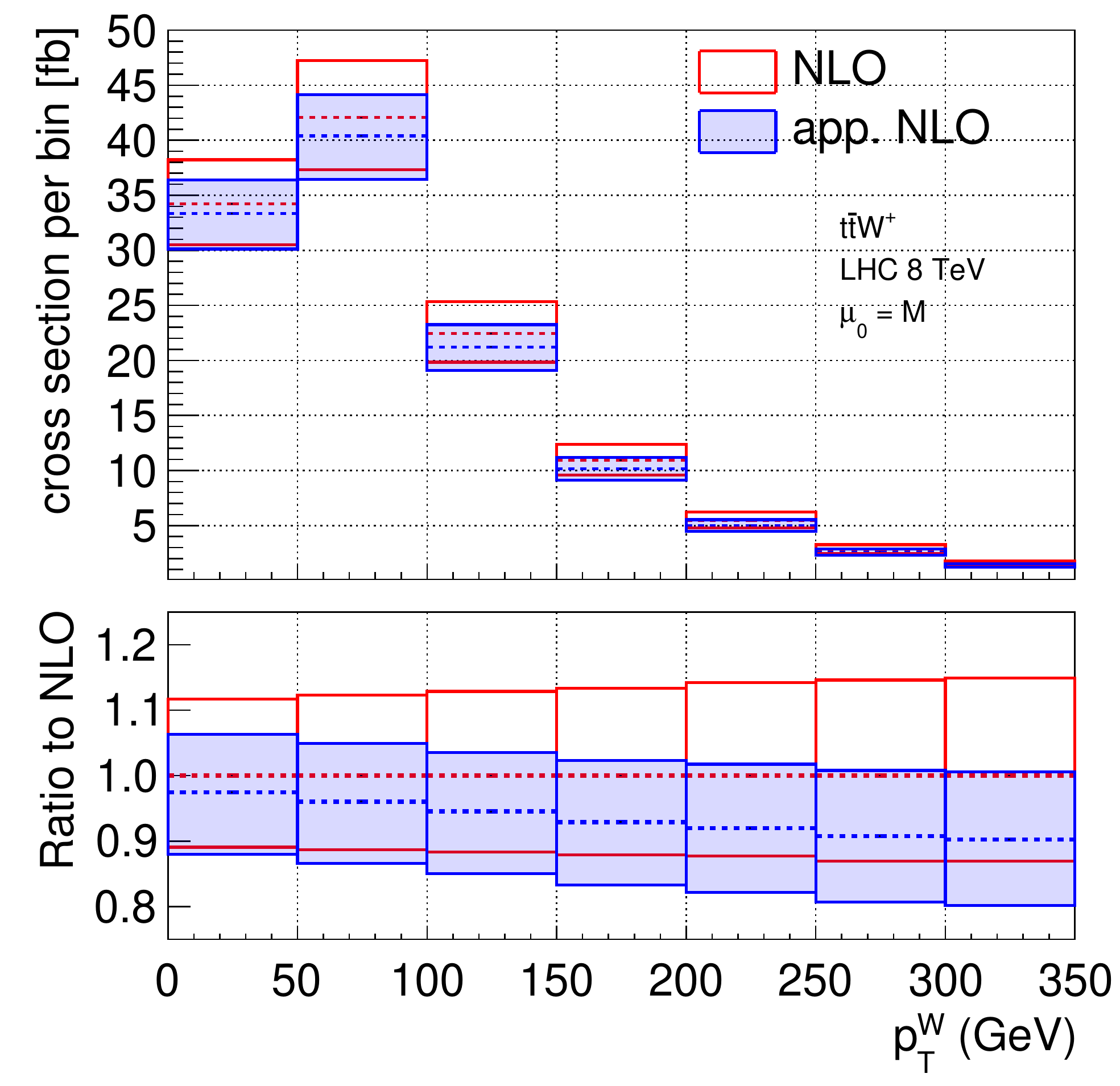} \\
		\end{tabular}
	\end{center}
	\caption{$t \bar{t} W^+$ production at $\sqrt{s} = 8$ TeV: Differential distributions at approximate NLO (blue band) compared to the complete NLO calculation carried out with \texttt{MG5} (red band).  The bands were obtained by varying the scale in the range $[M/2 ,2 M]$. MMHT 2014 NLO PDFs were used in all cases.
		\label{fig:Wp8nLOvsNLO}
	}
	
\end{figure}

\begin{figure}[tp]
	\begin{center}
		\begin{tabular}{cc}
			\includegraphics[width=7.2cm]{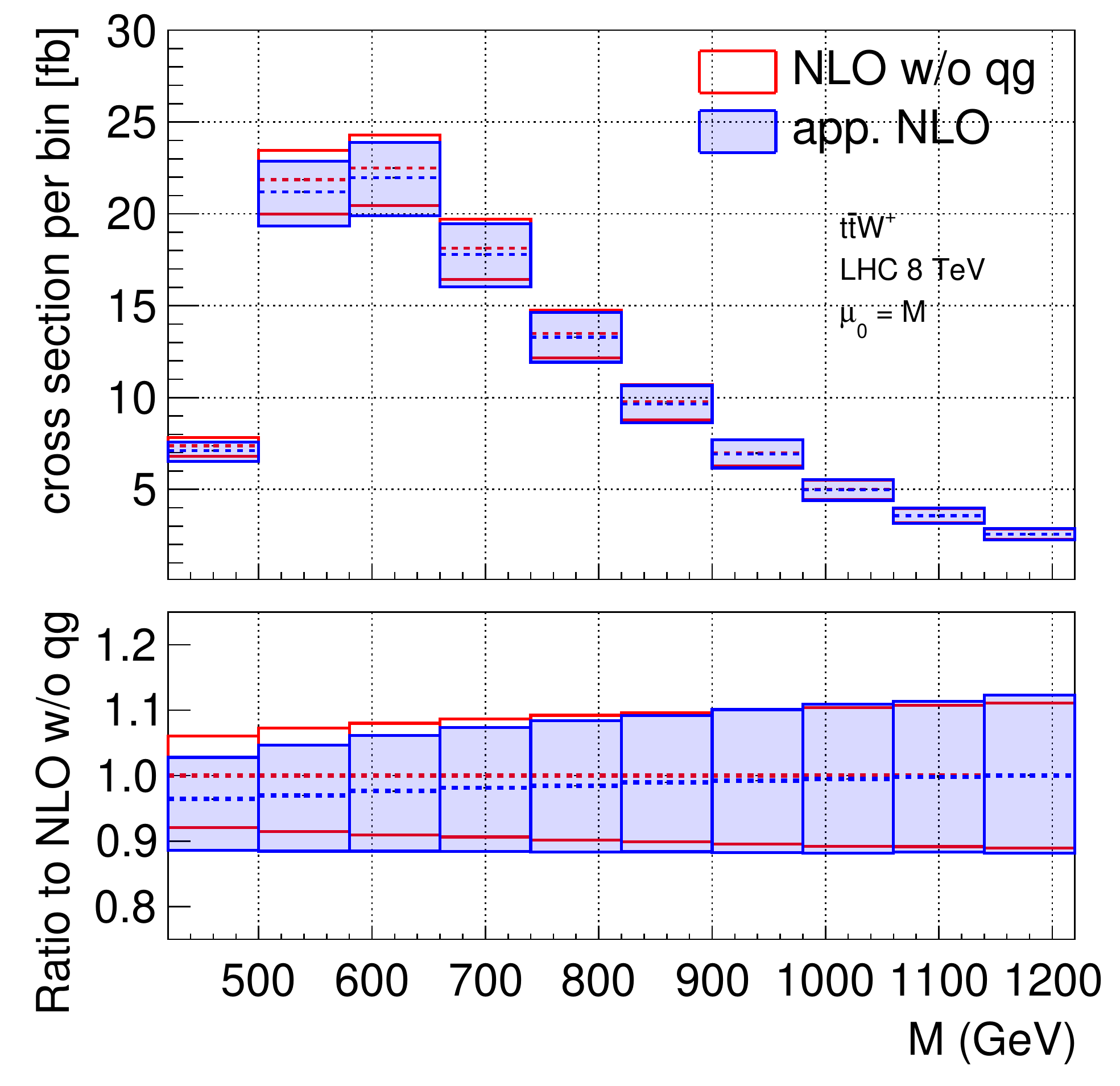} & \includegraphics[width=7.2cm]{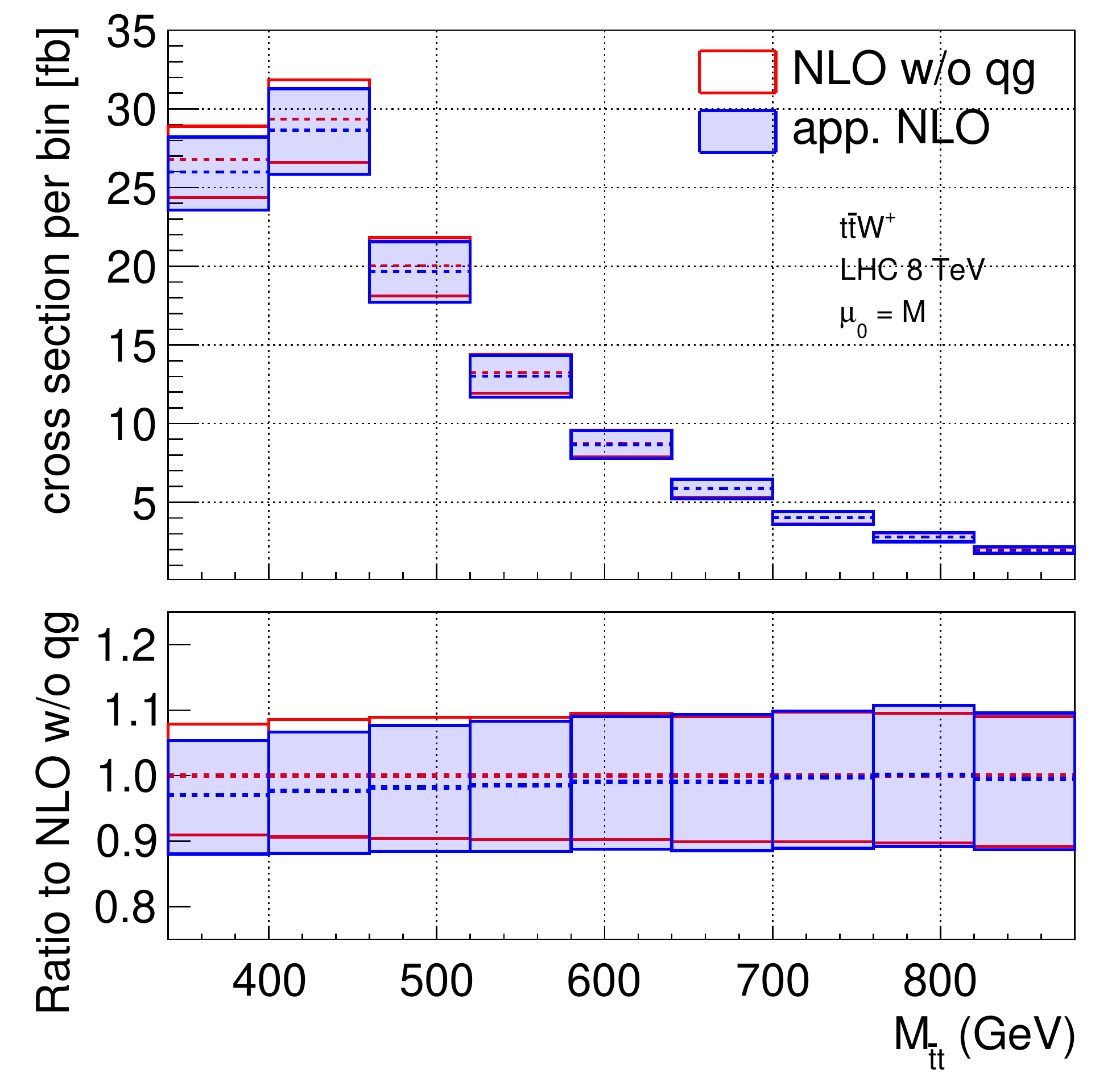} \\
			\includegraphics[width=7.2cm]{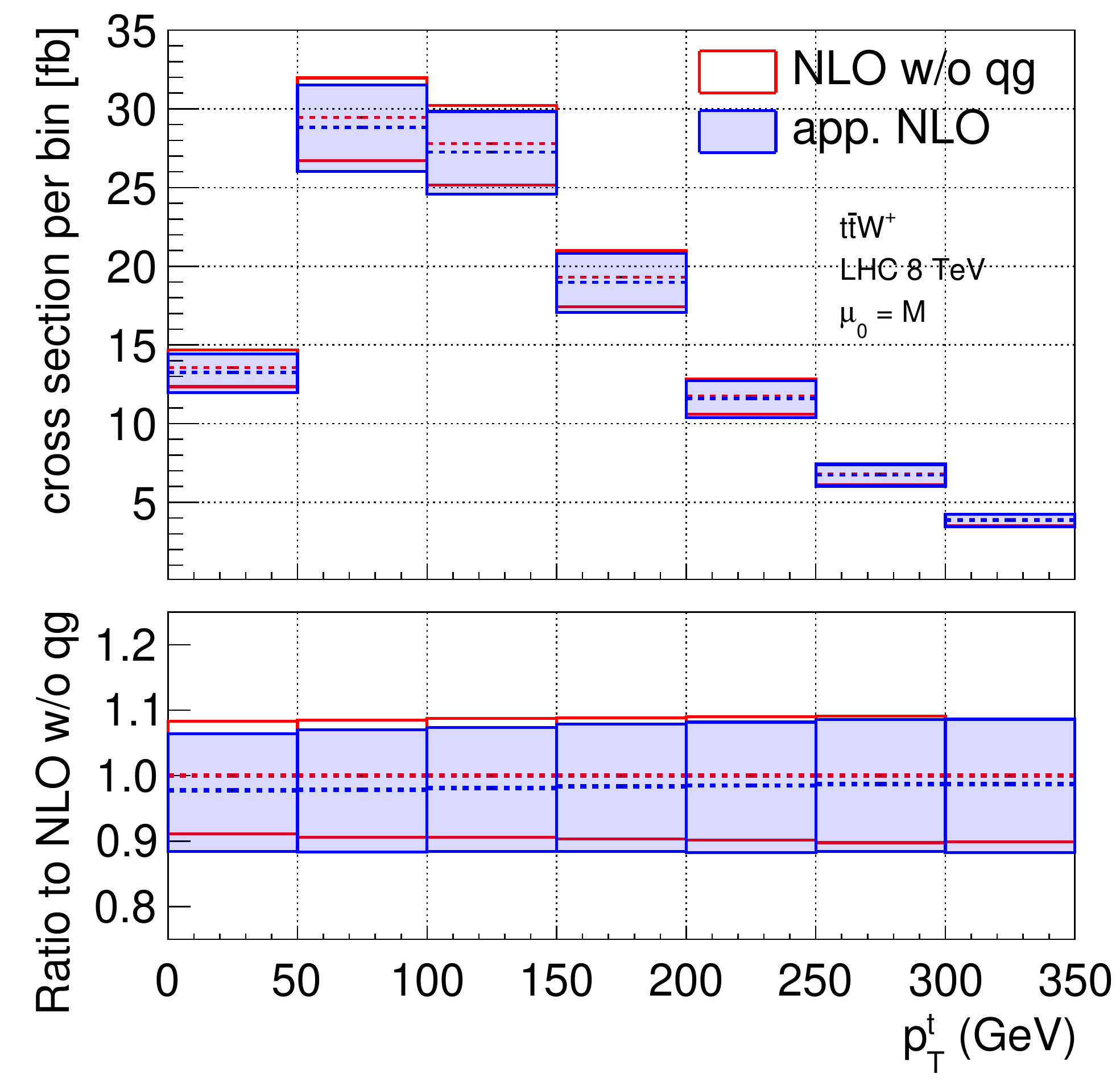} & \includegraphics[width=7.2cm]{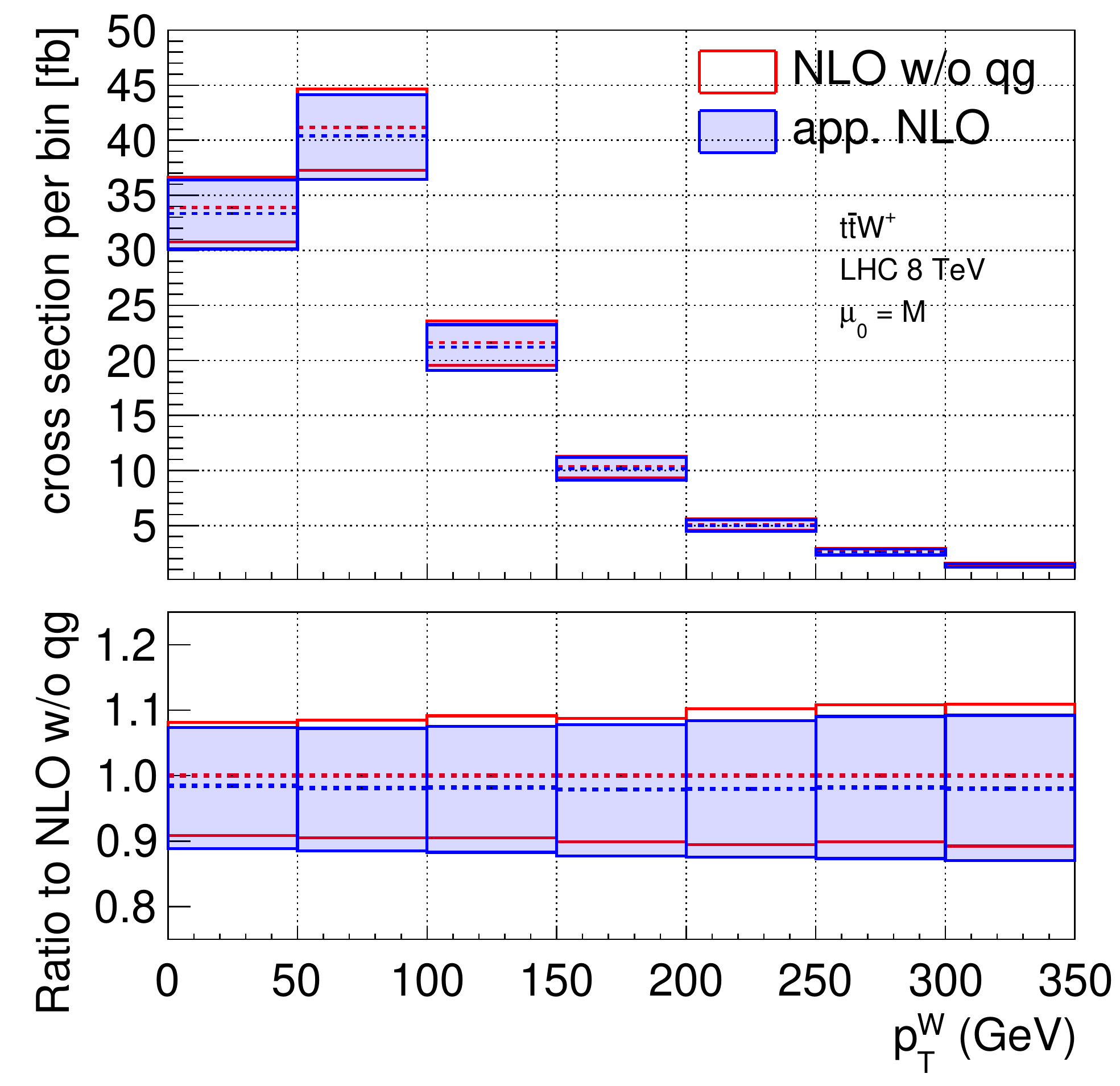} \\
		\end{tabular}
	\end{center}
	\caption{$t \bar{t} W^+$ production at $\sqrt{s} = 8$ TeV: Differential distributions at approximate NLO (blue band) compared to the NLO distributions without the quark-gluon channel contribution (red band). All settings are as in Figure~\ref{fig:Wp8nLOvsNLO}.
		\label{fig:Wp8nLOvsNLOnoqg}
	}
	
\end{figure}

We start by comparing the approximate NLO distributions with the complete NLO calculation of the same observables  carried out with \mgamc.
Figure~\ref{fig:Wp8nLOvsNLO} refers to $t \bar{t} W^+$ production at the LHC operating at a center of mass energy of $8$~TeV. The factorization/renormalization scale has been fixed to $M$ and varied, as usual, in the range $[M/2,2M]$. 
We see that the approximate NLO distributions (in blue)
are reasonably close to the full NLO distributions (in red), and have uncertainty bands that are smaller than but comparable to the ones found in NLO calculations. We also observe that the approximate NLO calculations tend to be slightly smaller than the complete NLO ones. However, in each panel the lower plot shows that the ratio of the approximate NLO over the full NLO (evaluated at $\mu_f = M$), represented by the  histogram in blue, is not completely flat. 
On the contrary, by looking at  Figure~\ref{fig:Wp8nLOvsNLOnoqg} one can see that if one excludes the contribution of the quark-gluon channel from the NLO distributions, the approximate NLO calculation reproduces quite well also the shape of the distributions: The approximate NLO result is very close to the NLO one (without the quark-gluon channel contribution) in each bin. Figure~\ref{fig:Wp8nLOvsNLOnoqg} also shows that the scale uncertainty bands at approximate NLO are almost identical to the NLO bands without the quark gluon channel.

\begin{figure}[tp]
	\begin{center}
		\begin{tabular}{cc}
			\includegraphics[width=7cm]{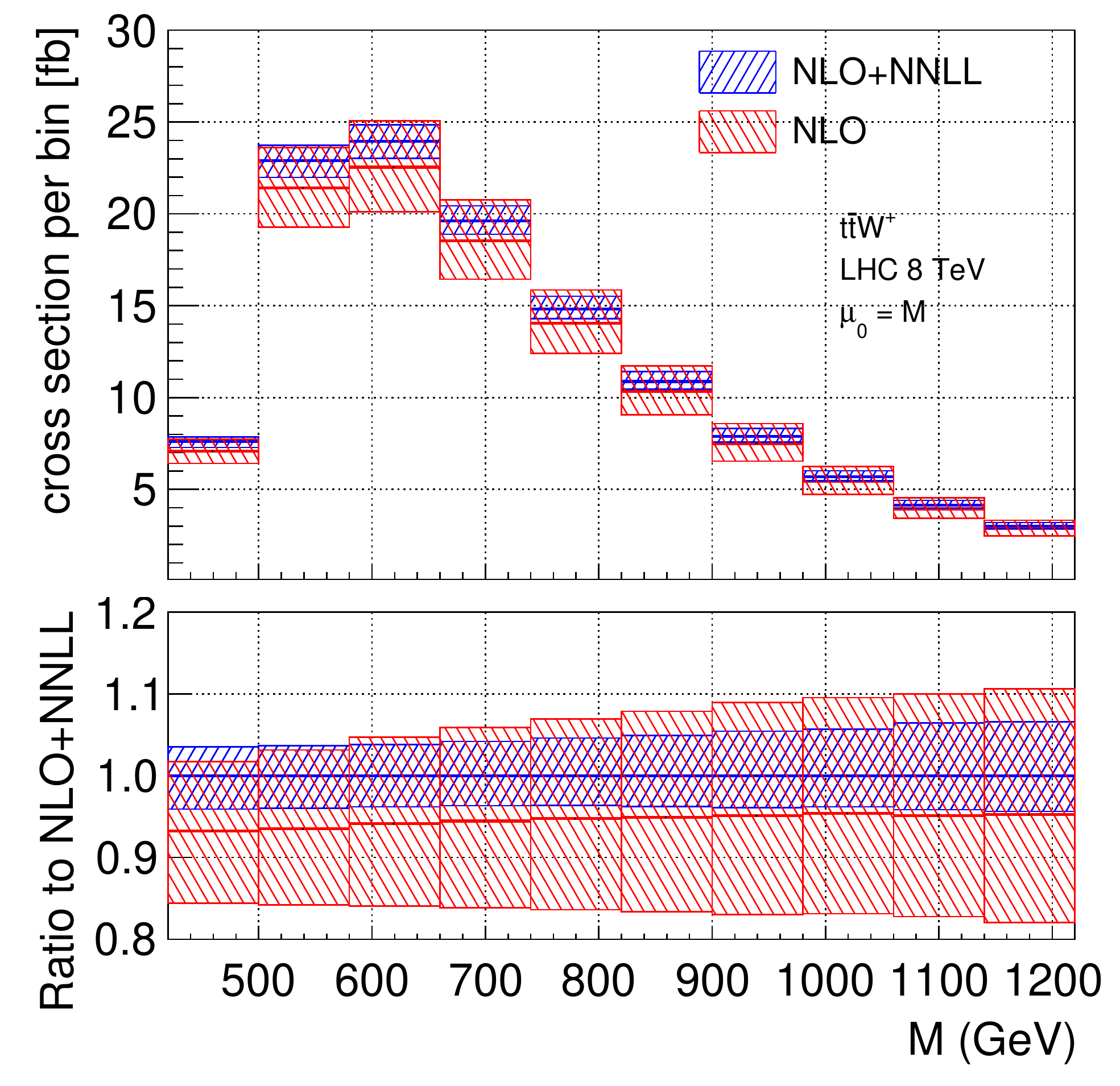} & \includegraphics[width=7cm]{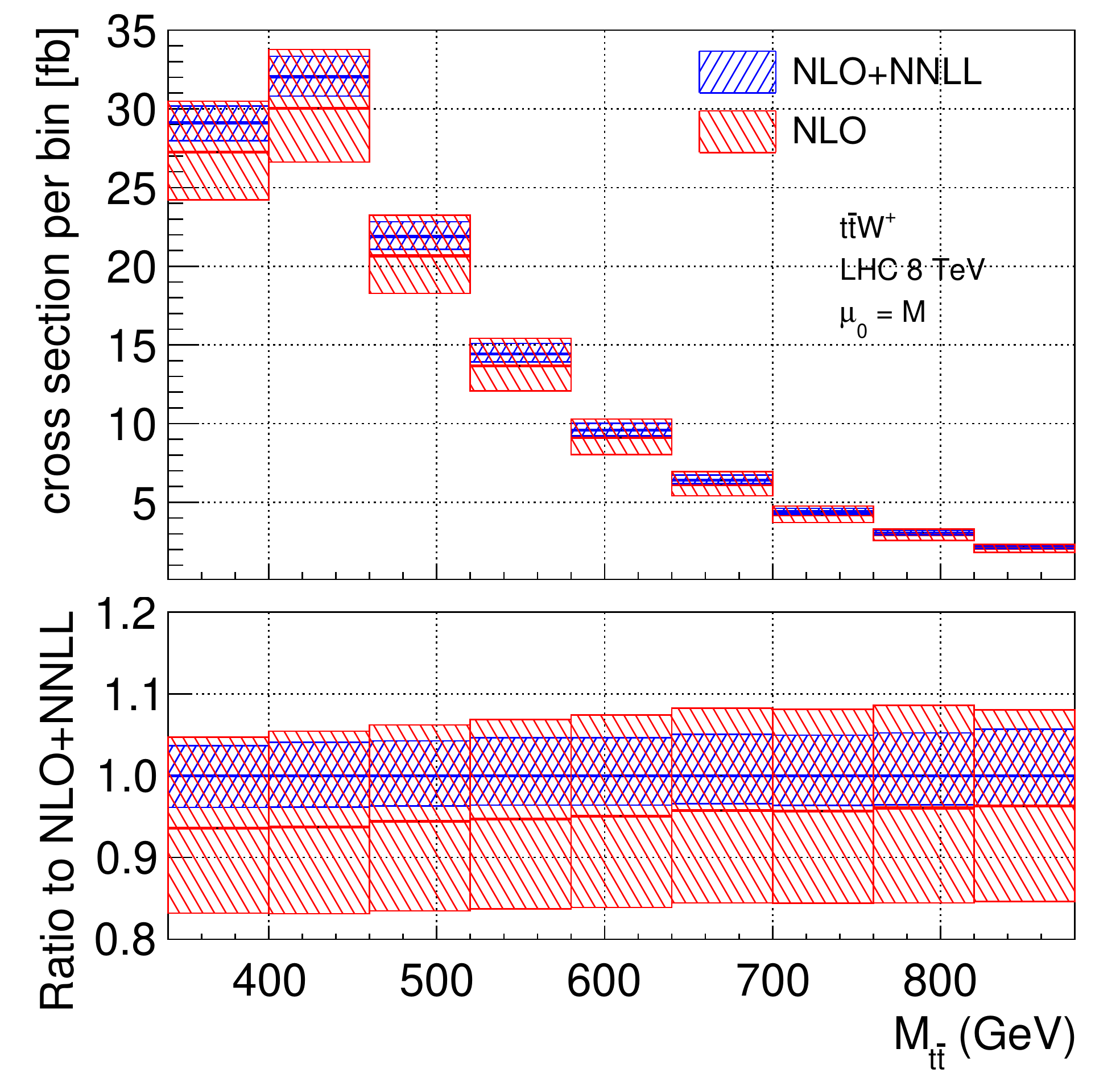} \\
			\includegraphics[width=7cm]{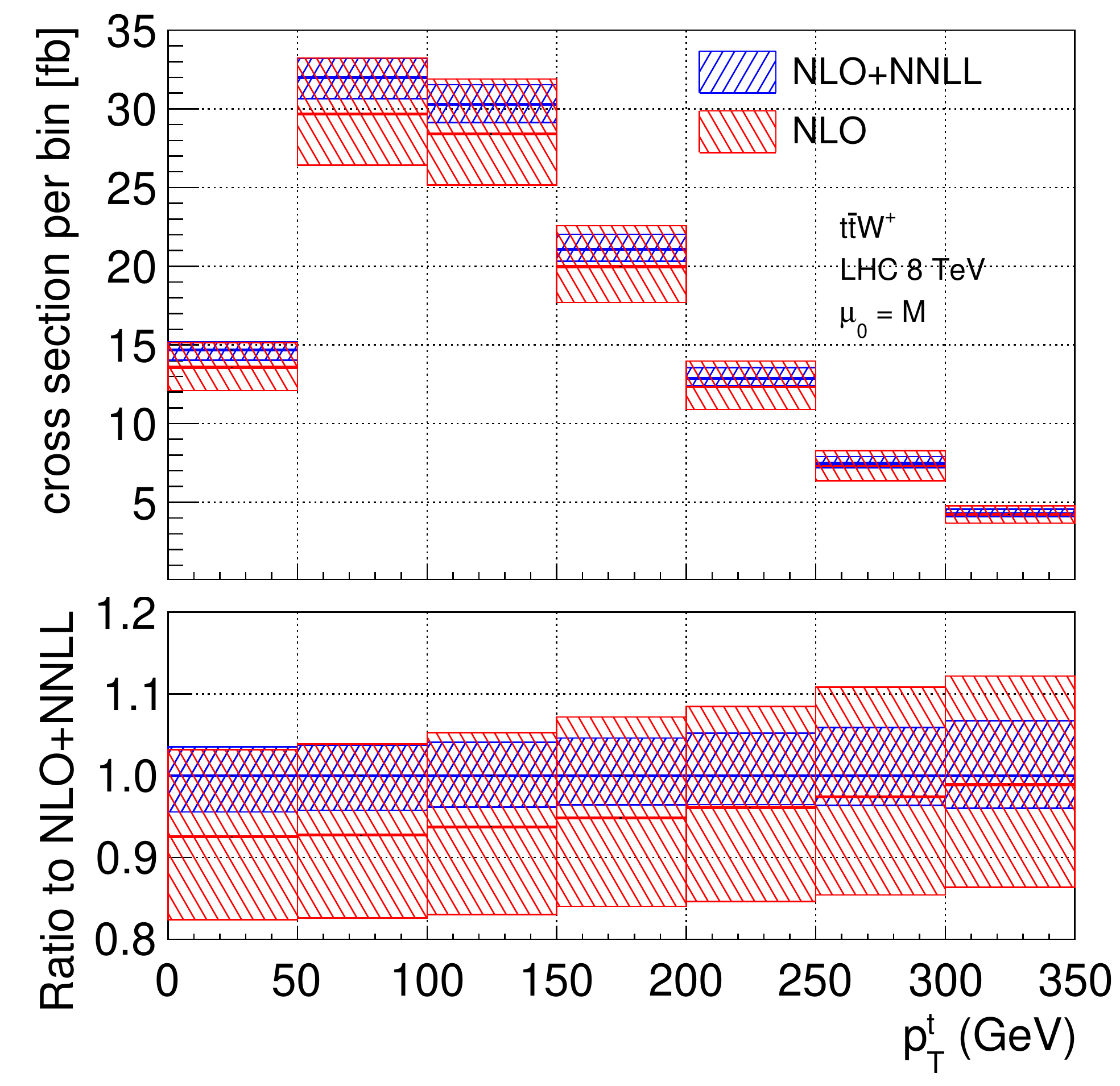} & \includegraphics[width=7cm]{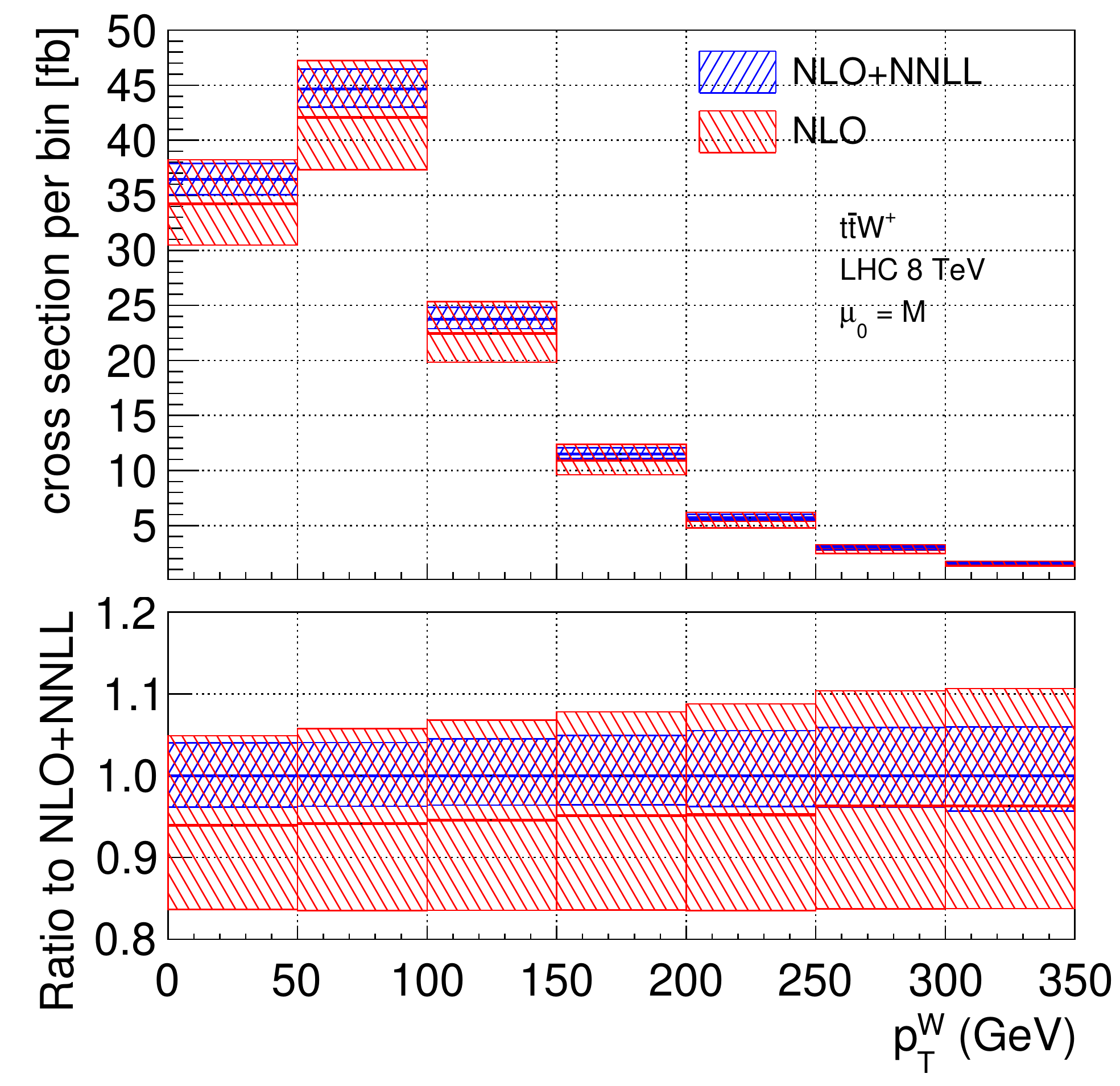} \\
		\end{tabular}
	\end{center}
	\caption{$t \bar{t} W^+$ production at $\sqrt{s} = 8$ TeV: Differential distributions at NLO+NNLL (blue band) compared to the NLO calculation (red band).
		 MMHT 2014 NNLO PDFs were used for the NLO+NNLL calculation, while the NLO calculation was carried out with  MMHT 2014 NLO PDFs. 
		\label{fig:Wp8NLOvsNNLL}
	}
\end{figure}

\begin{figure}[tp]
	\begin{center}
		\begin{tabular}{cc}
			\includegraphics[width=7cm]{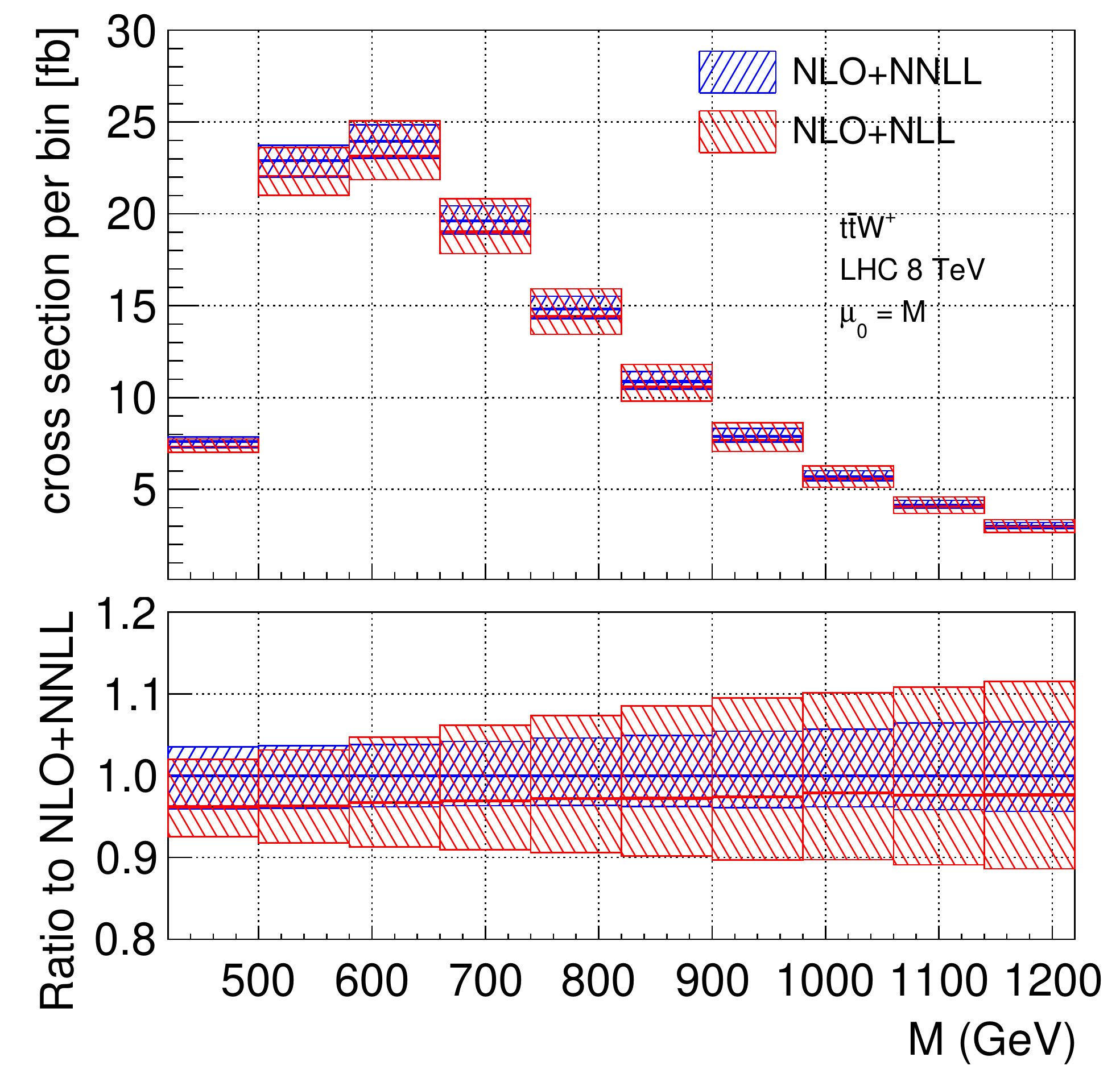} & \includegraphics[width=7cm]{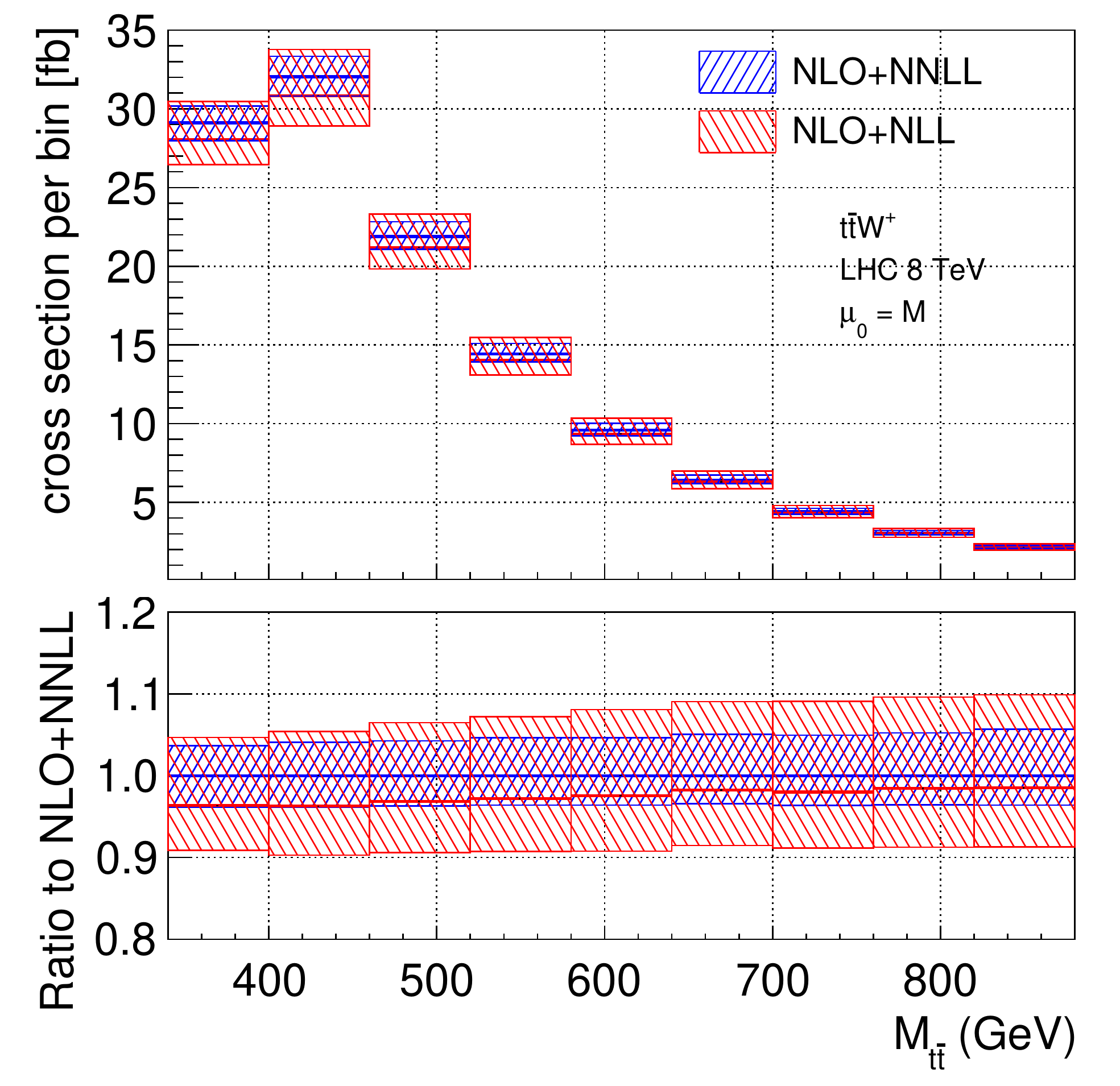} \\
			\includegraphics[width=7cm]{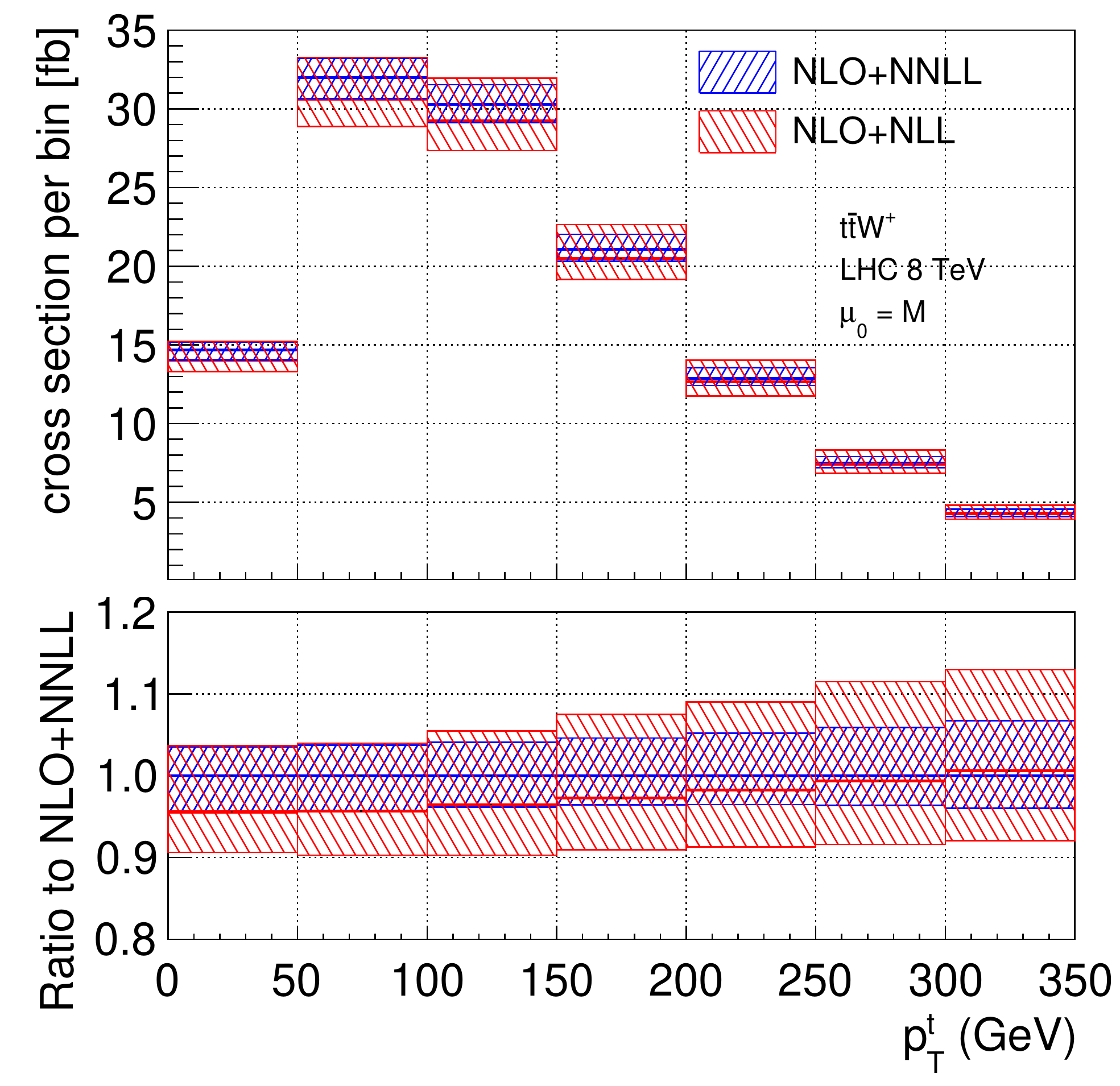} & \includegraphics[width=7cm]{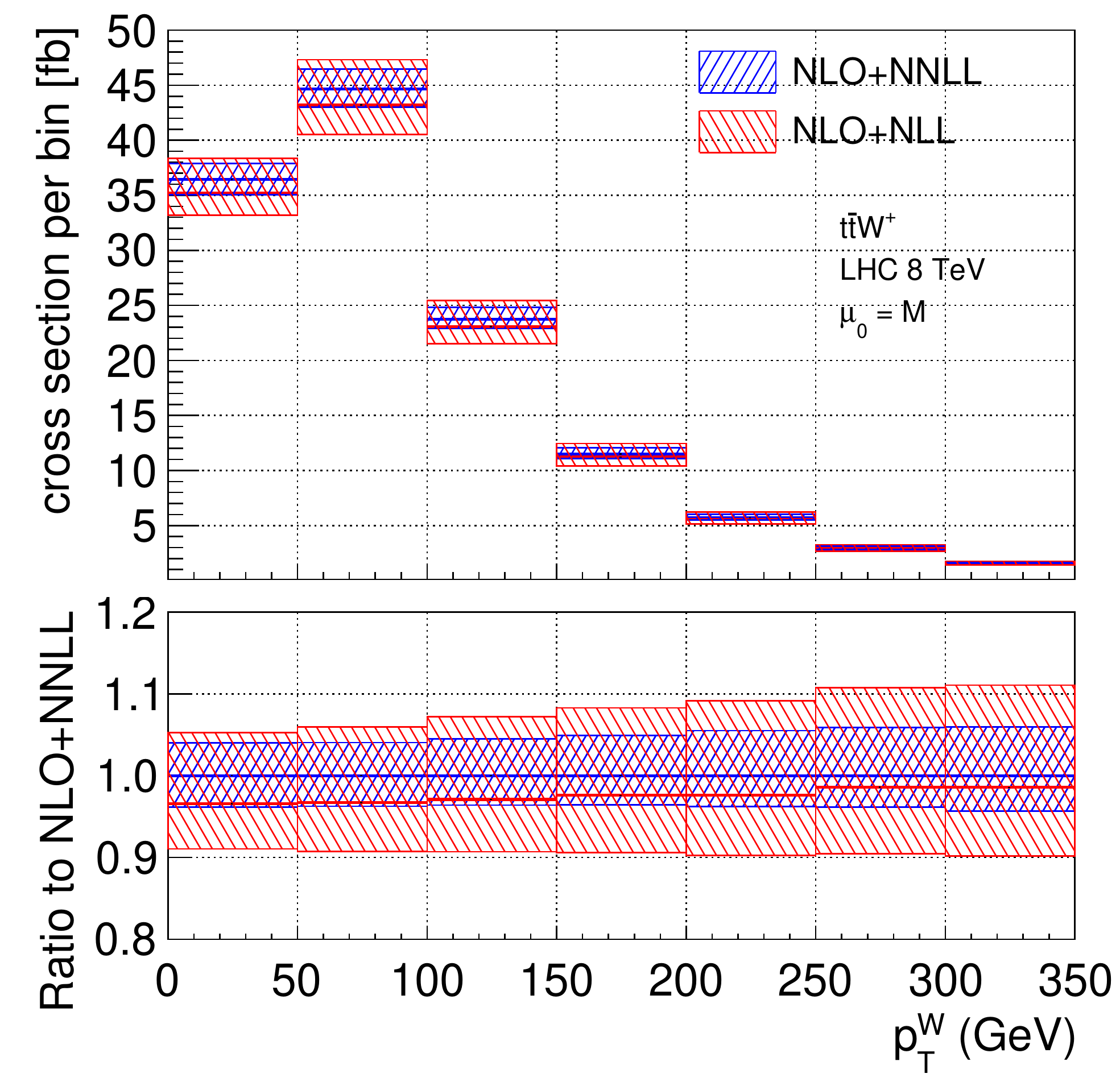} \\
		\end{tabular}
	\end{center}
	\caption{$t \bar{t} W^+$ production at $\sqrt{s} = 8$ TeV: Differential distributions at NLO+NNLL (blue band) compared to the NLO+NLL calculation (red band). MMHT 2014 NNLO PDFs were used for the NLO+NNLL calculation, while the NLO+NLL calculation was carried out with  MMHT 2014 NLO PDFs. 
		\label{fig:Wp8NLLvsNNLL}
	}
\end{figure}

\begin{figure}[tp]
	\begin{center}
		\begin{tabular}{cc}
			\includegraphics[width=7cm]{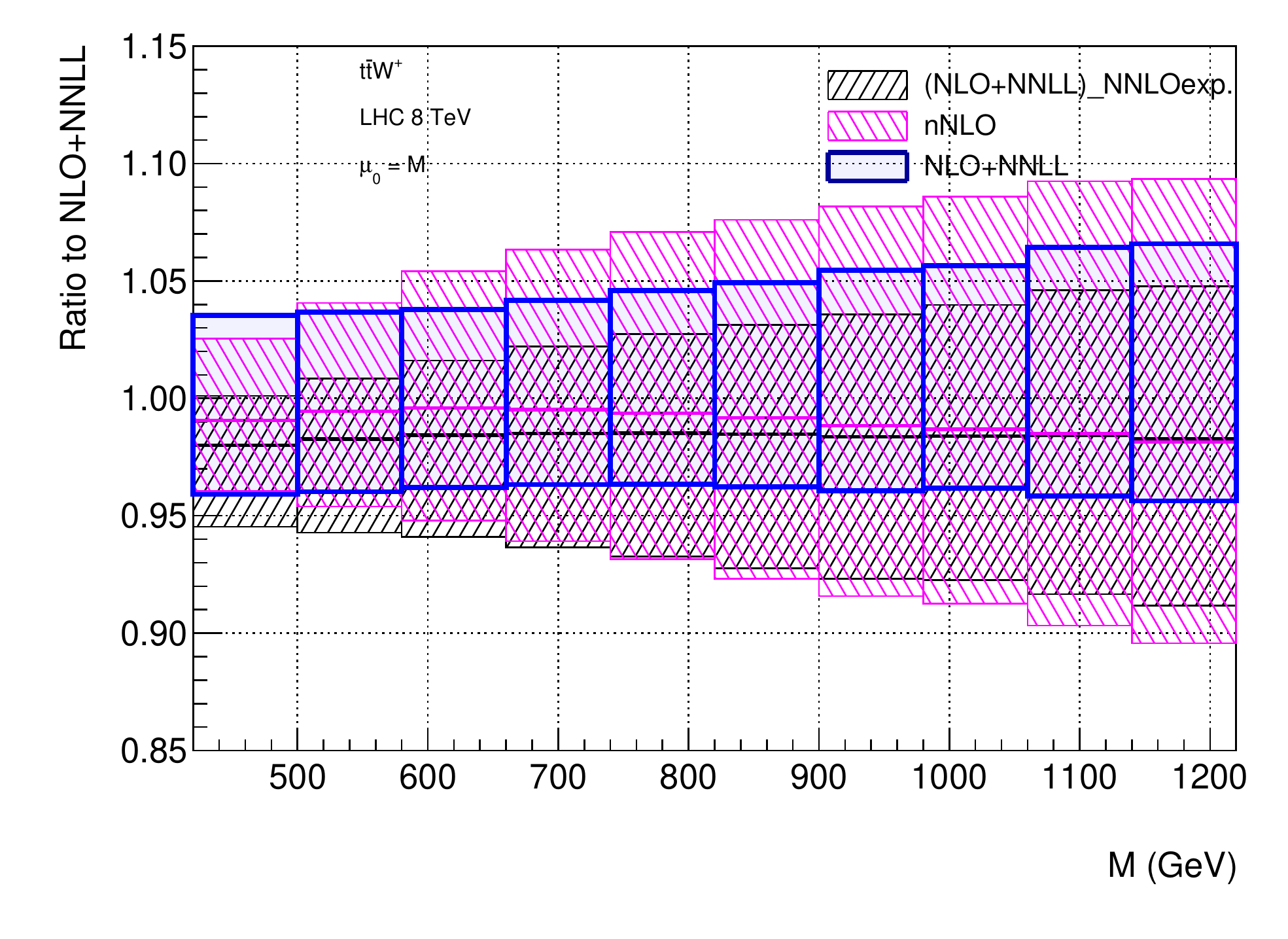} & \includegraphics[width=7cm]{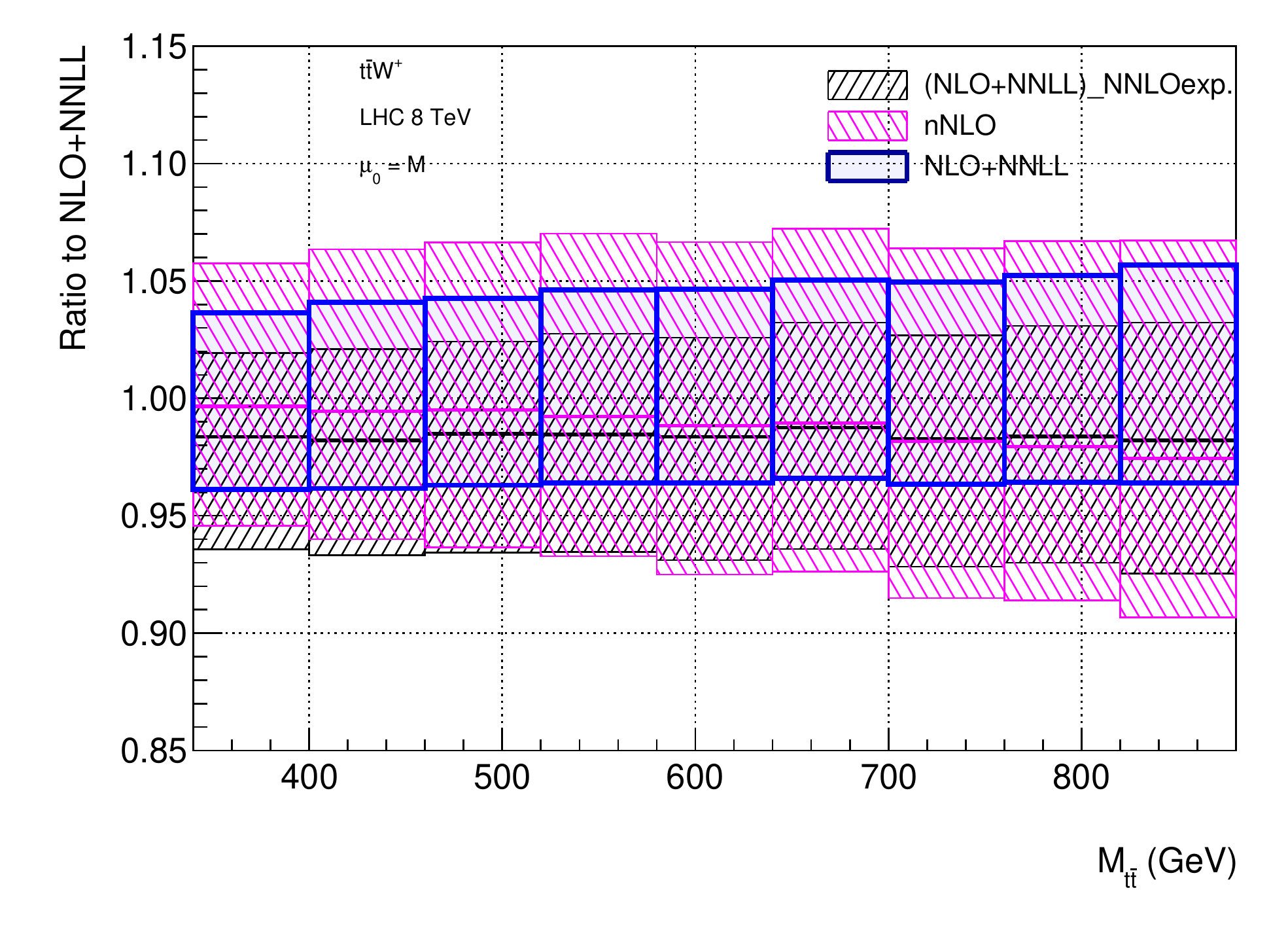} \\
			\includegraphics[width=7cm]{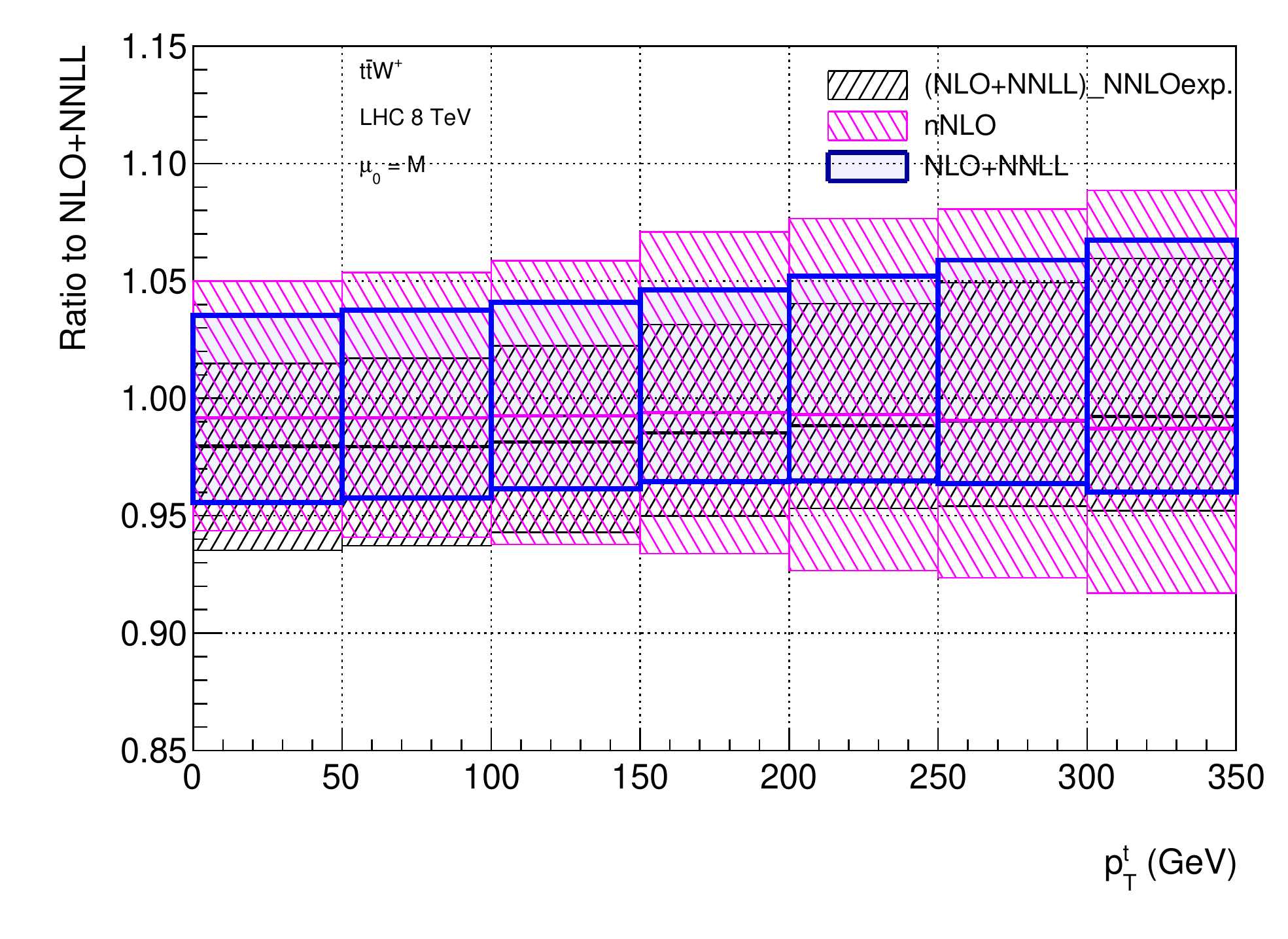} & \includegraphics[width=7cm]{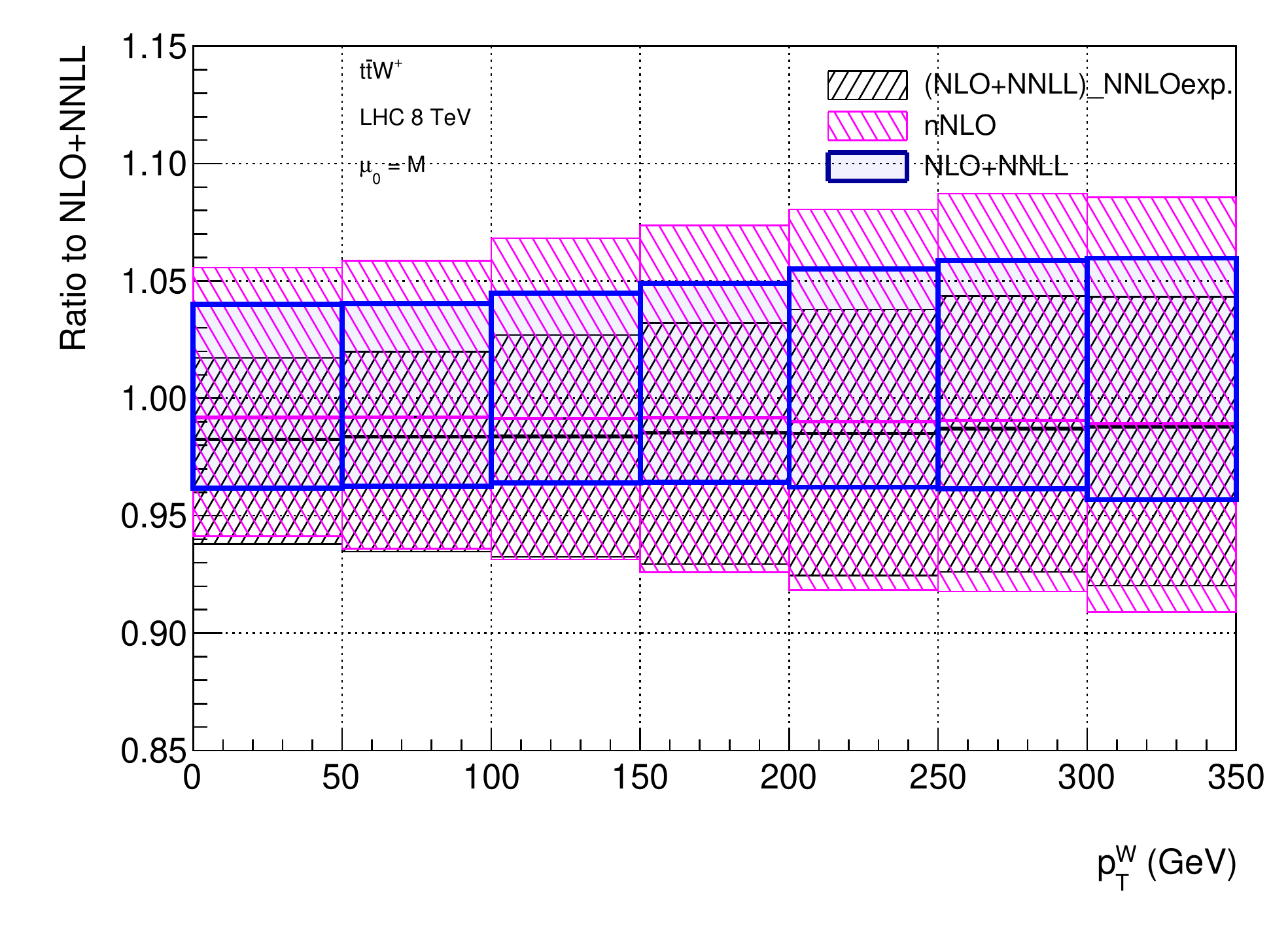} \\
		\end{tabular}
	\end{center}
	\caption{$t \bar{t} W^+$ production at $\sqrt{s} = 8$ TeV: Differential distributions ratios.  MMHT 2014 NNLO PDFs were used in all cases. 
		\label{fig:Wp8NNLOrats}
	}
\end{figure}

\begin{figure}[tp]
	\begin{center}
		\begin{tabular}{cc}
			\includegraphics[width=7.2cm]{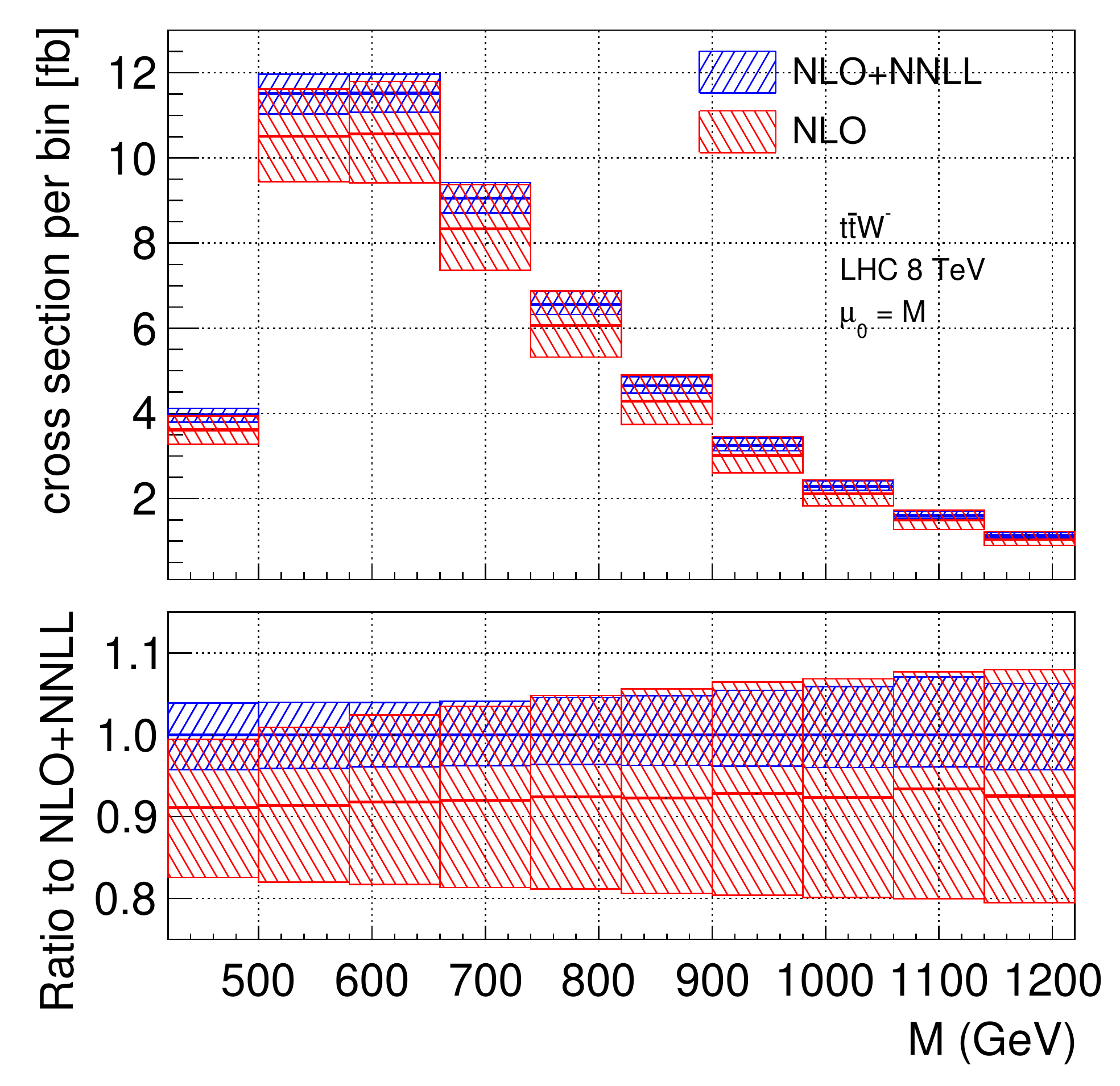} & \includegraphics[width=7.2cm]{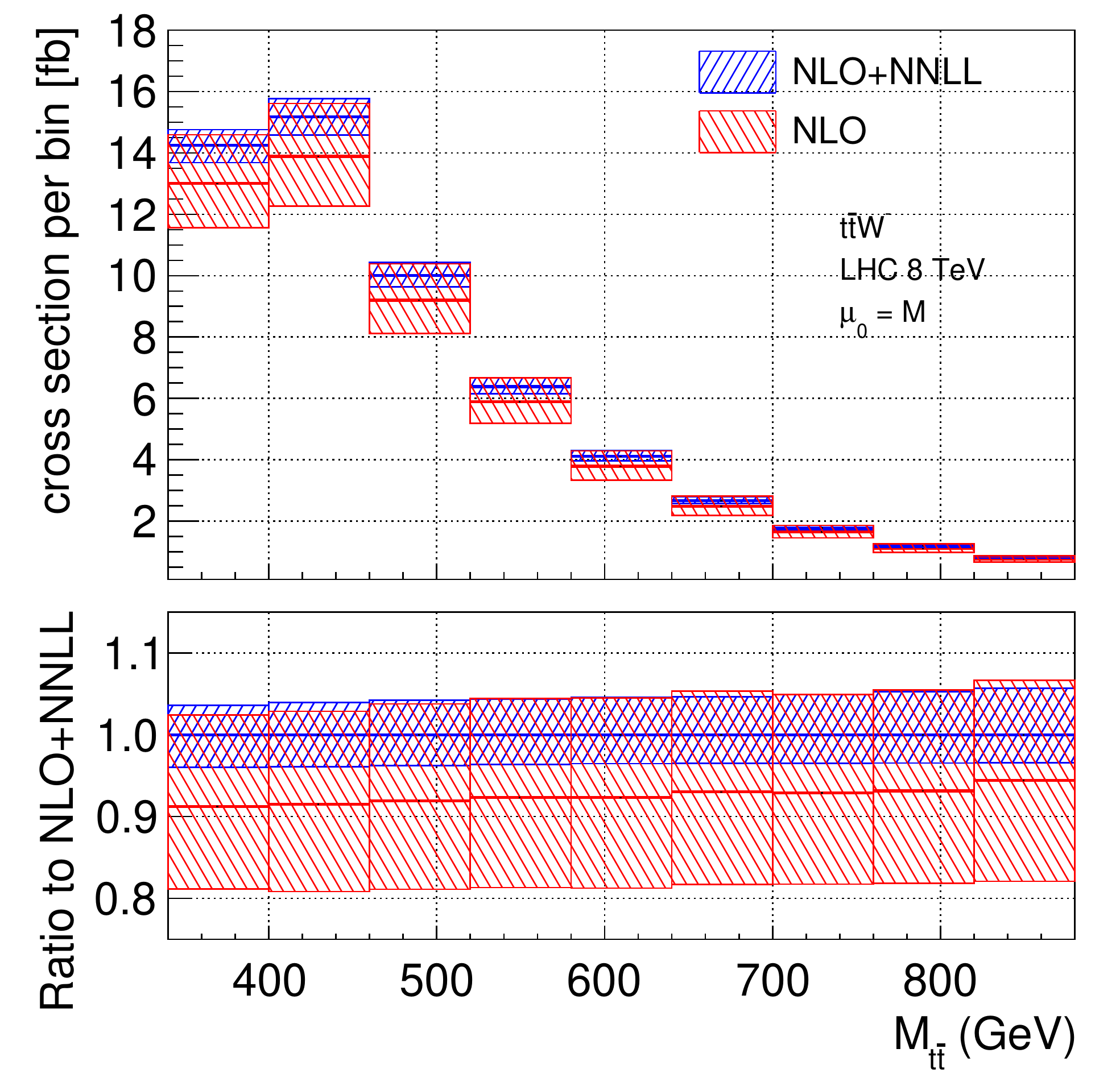} \\
			\includegraphics[width=7.2cm]{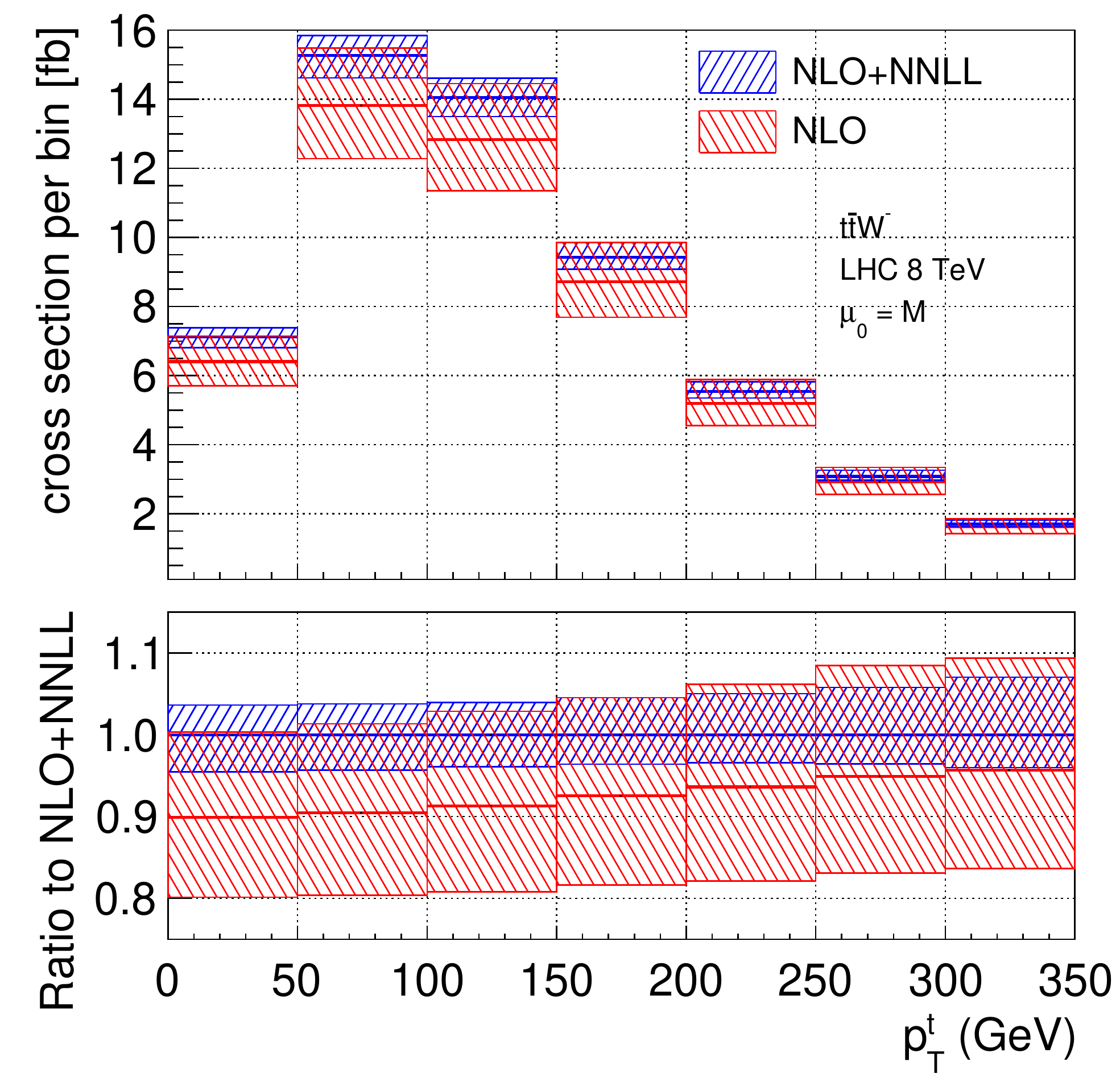} & \includegraphics[width=7.2cm]{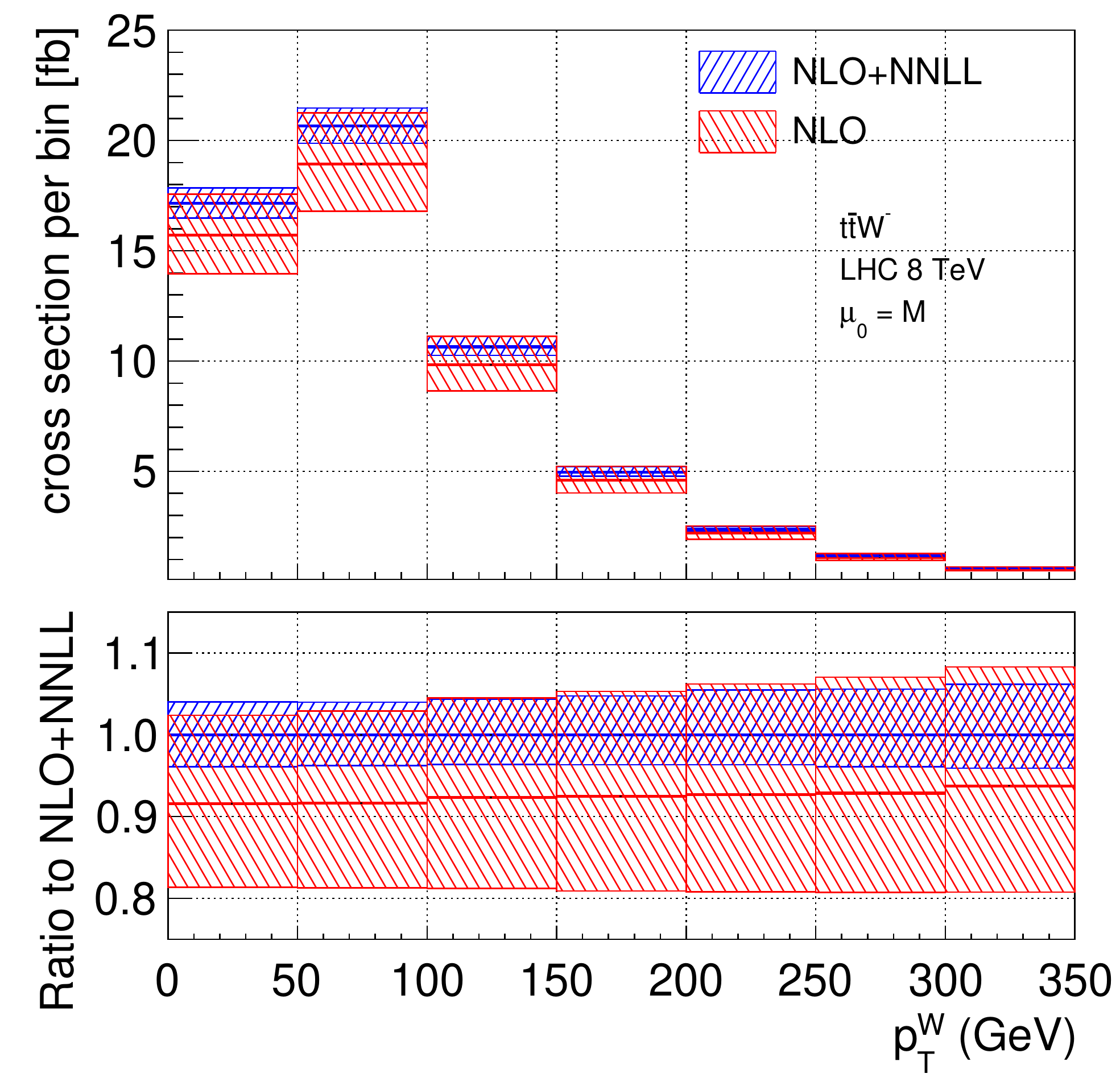} \\
		\end{tabular}
	\end{center}
	\caption{$t \bar{t} W^-$ production  at $\sqrt{s} = 8$ TeV: Differential distributions at NLO+NNLL (blue bands) compared to the  NLO calculation (red bands). MMHT 2014 NNLO PDFs were used for the NLO+NNLL calculation, while the NLO calculation was carried out with  MMHT 2014 NLO PDFs. 
		\label{fig:Wm8NNLL}
	}
\end{figure}

Keeping in mind that the NLO contribution of the quark-gluon channel is included in the NLO+NNLL calculations through the matching procedure, we consider now the predictions for the differential distributions at NLO+NNLL.


Figure~\ref{fig:Wp8NLOvsNNLL} refers to $t \bar{t} W^+$ production at the LHC operating at a center of mass energy of $8$~TeV  and compares the NLO+NNLL distributions (blue bands) to the NLO distributions (red bands). The ratio plots in each panel were obtained by dividing each bin by the NLO+NNLL predictions evaluated with setting all scales in the resummed calculation to their default values. One can observe that the NLO+NNLL bands have a significant overlap with the upper part of the NLO bands. The width of the NLO+NNLL bands is roughly half of the width of the NLO bands or smaller in almost all distributions and bins shown in  Figure~\ref{fig:Wp8NLOvsNNLL}.
 

Figure~\ref{fig:Wp8NLLvsNNLL} compares the NLO+NNLL distributions (blue bands) to the 
corresponding NLO+NLL distributions (red bands). NLO+NNLL calculations give slightly larger results than the NLO+NLL ones. The NLO+NNLL uncertainty bands are narrower than the NLO+NLL bands in each bin, and in particular in the tail of the distributions.
This shows that while NLO+NLL is already an improvement over NLO, the higher-order resummation effects contained in the NLO+NNLL result are not insignificant.

Finally we conclude our analysis of $t \bar{t} W^+$ production at $\sqrt{s} =  8$~TeV
by comparing NLO+NNLL, nNLO and NLO+NNLL expanded predictions in Figure~\ref{fig:Wp8NNLOrats}. This figure shows the ratio, separately for each bin, of the distribution to the NLO+NNLL calculation evaluated with default choices of all scales.
The blue band refers to NLO+NNLL calculations, the dashed red band to nNLO calculations and the dashed black band to the expansion of the NLO+NNLL resummation formula to order $\alpha_s^2$ relative to the leading order (which is of order $\alpha_s^2 \alpha$).  One can observe that the NLO+NNLL band is narrower than the nNLO one in particular in the tail of the distributions. The comparison of the NLO+NNLL band to the NLO+NNLL expanded band shows the impact of the terms of relative order $\alpha_s^3$ and higher, which are included in the NLO+NNLL calculation but are excluded from the NLO+NNLL expanded one. One can observe that these terms have the effect of increasing slightly the distributions bands in all bins.
We thus conclude that, in contrast to the case of of $t\bar{t}$ production \cite{Pecjak:2016nee, Ferroglia:2015ivv}, the beyond NNLO resummation effects are relatively moderate even in the high-energy tails of the distributions.

\begin{figure}[tp]
	\begin{center}
		\begin{tabular}{cc}
			\includegraphics[width=7.2cm]{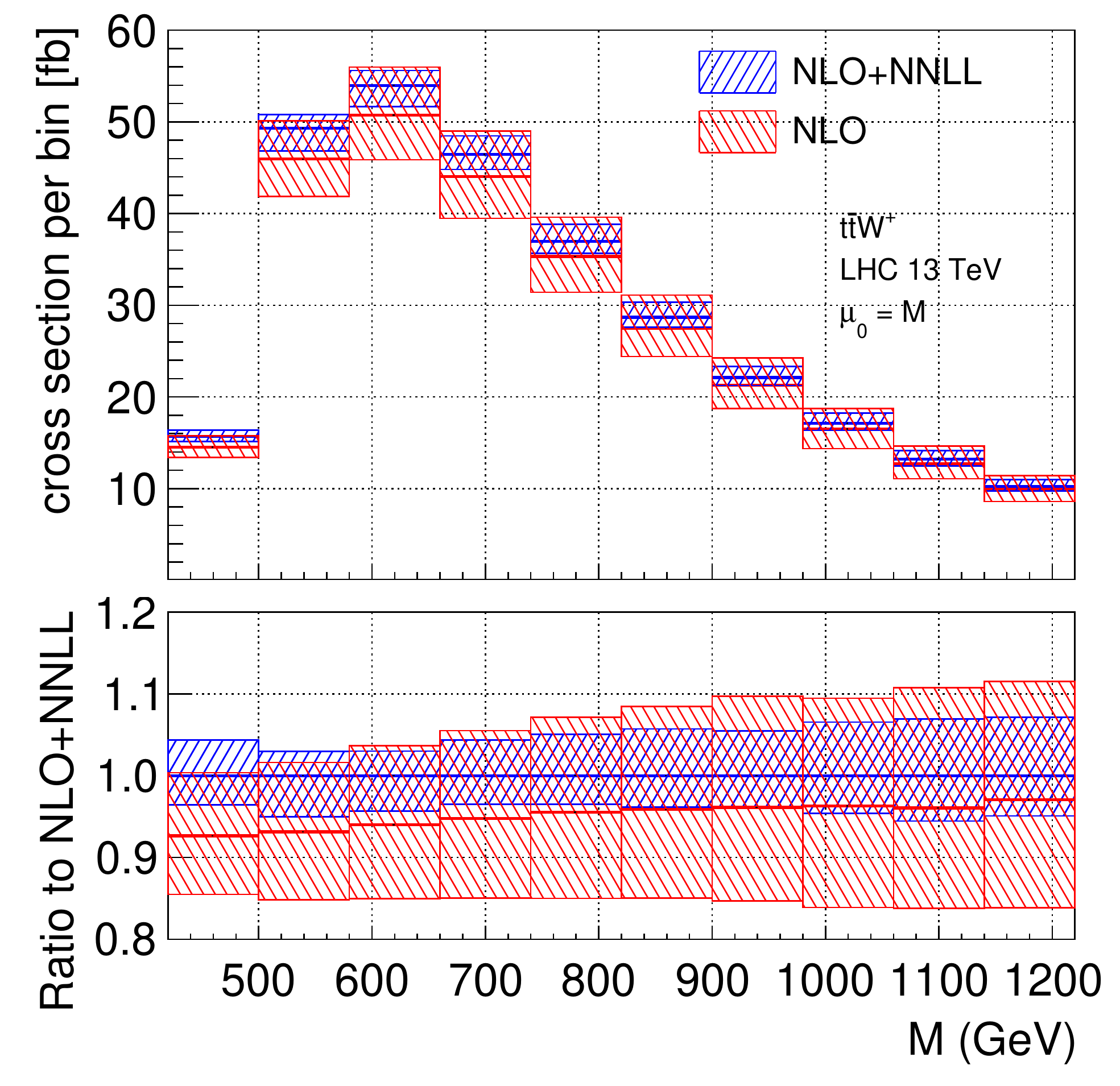} & \includegraphics[width=7.2cm]{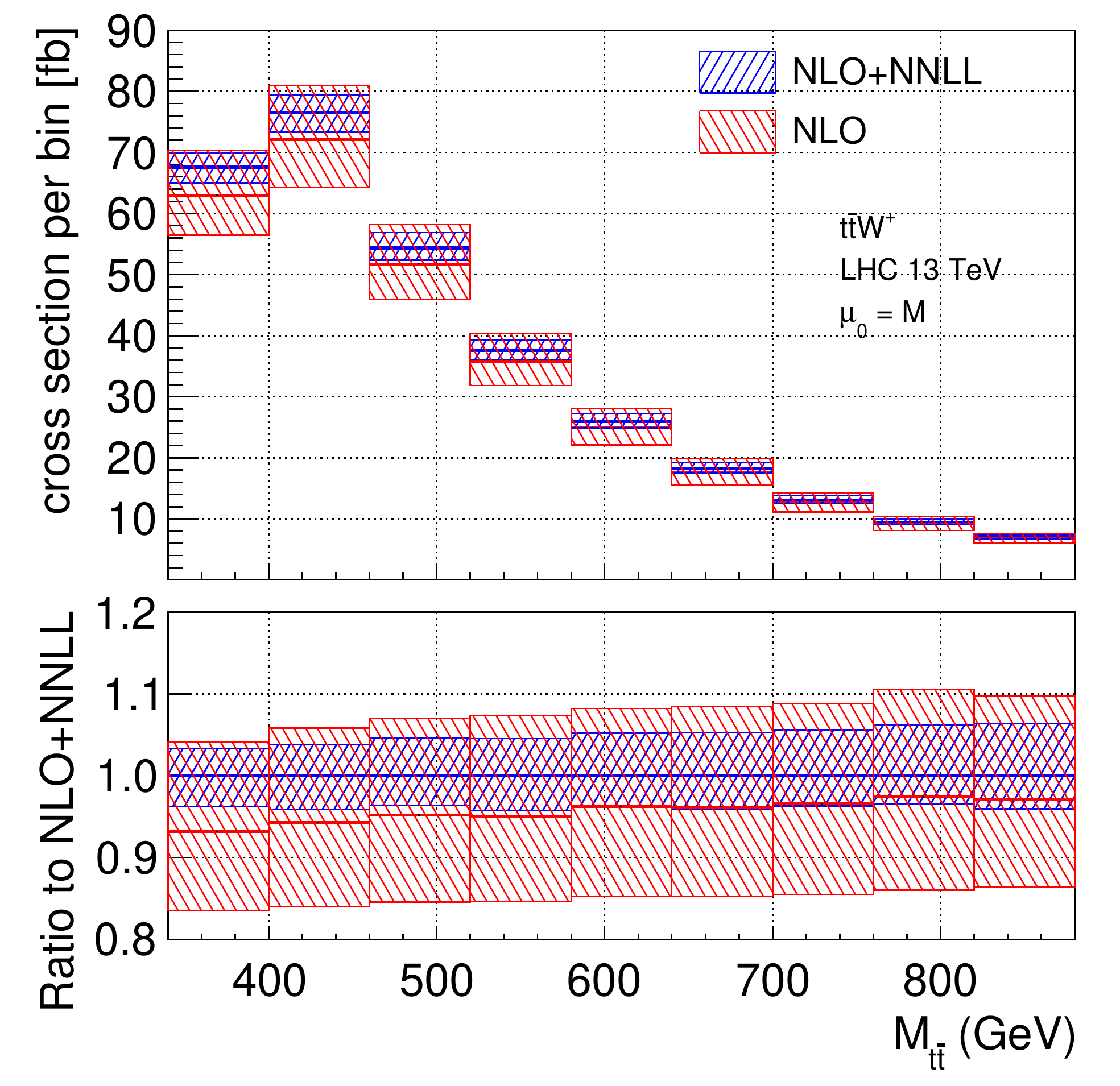} \\
			\includegraphics[width=7.2cm]{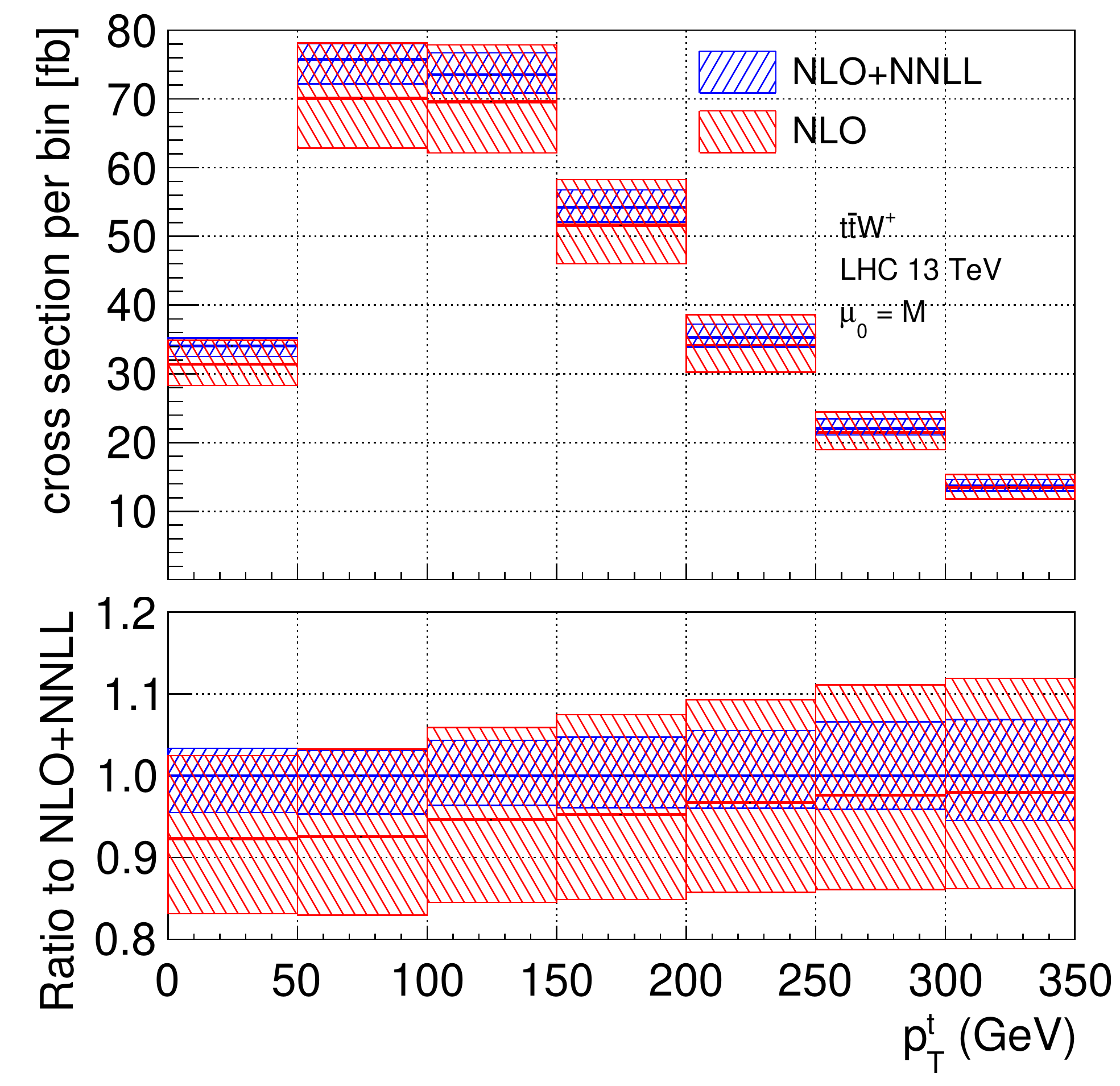} & \includegraphics[width=7.2cm]{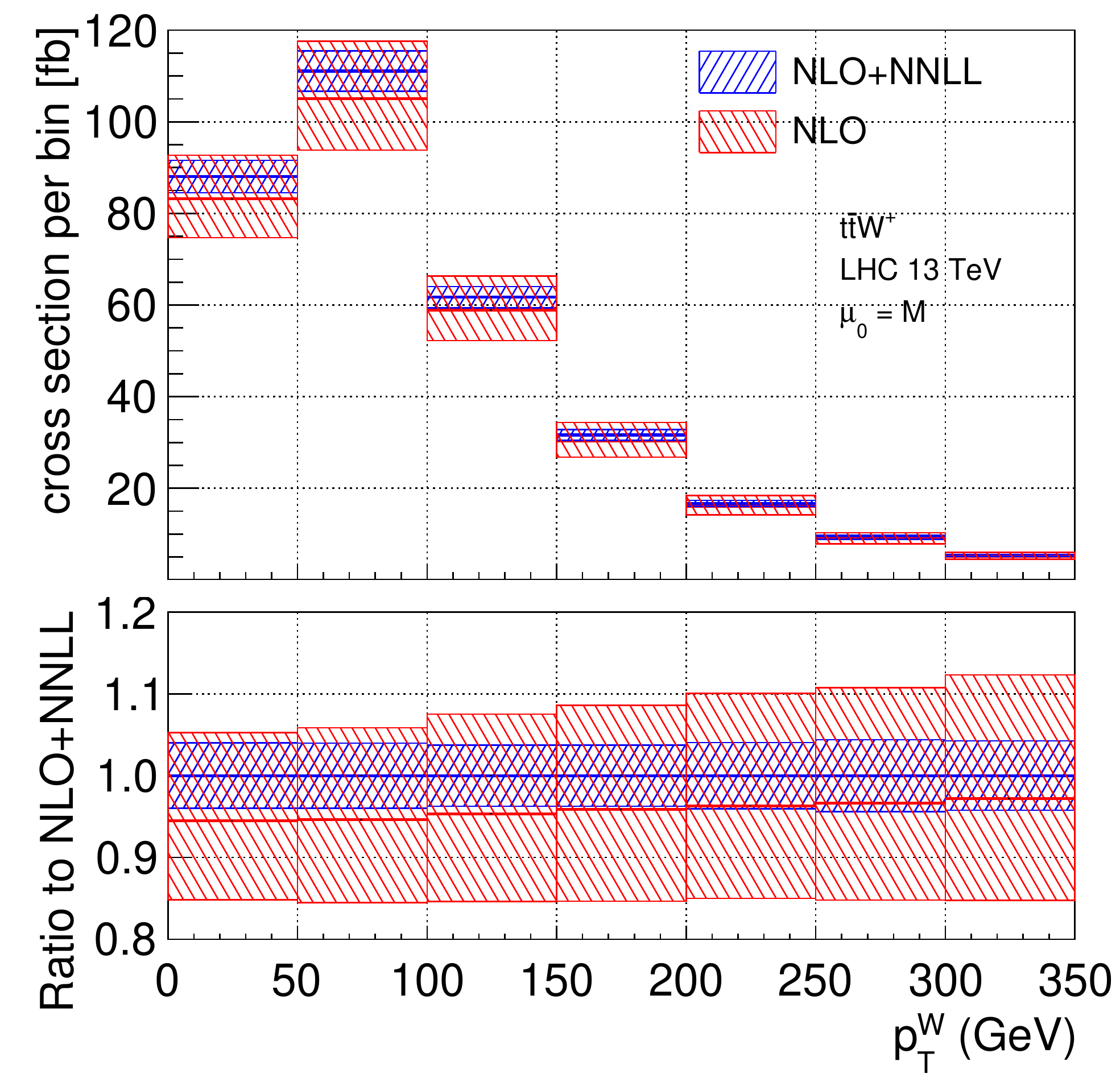} \\
		\end{tabular}
	\end{center}
	\caption{$t \bar{t} W^+$ production at $\sqrt{s} = 13$ TeV: Differential distributions at NLO+NNLL (blue bands) compared to the  NLO calculation (red bands). MMHT 2014 NNLO PDFs were used for the NLO+NNLL calculation, while the NLO calculation was carried out with  MMHT 2014 NLO PDFs. 
		\label{fig:Wp13NNLL}
	}
\end{figure}

\begin{figure}[tp]
	\begin{center}
		\begin{tabular}{cc}
			\includegraphics[width=7.2cm]{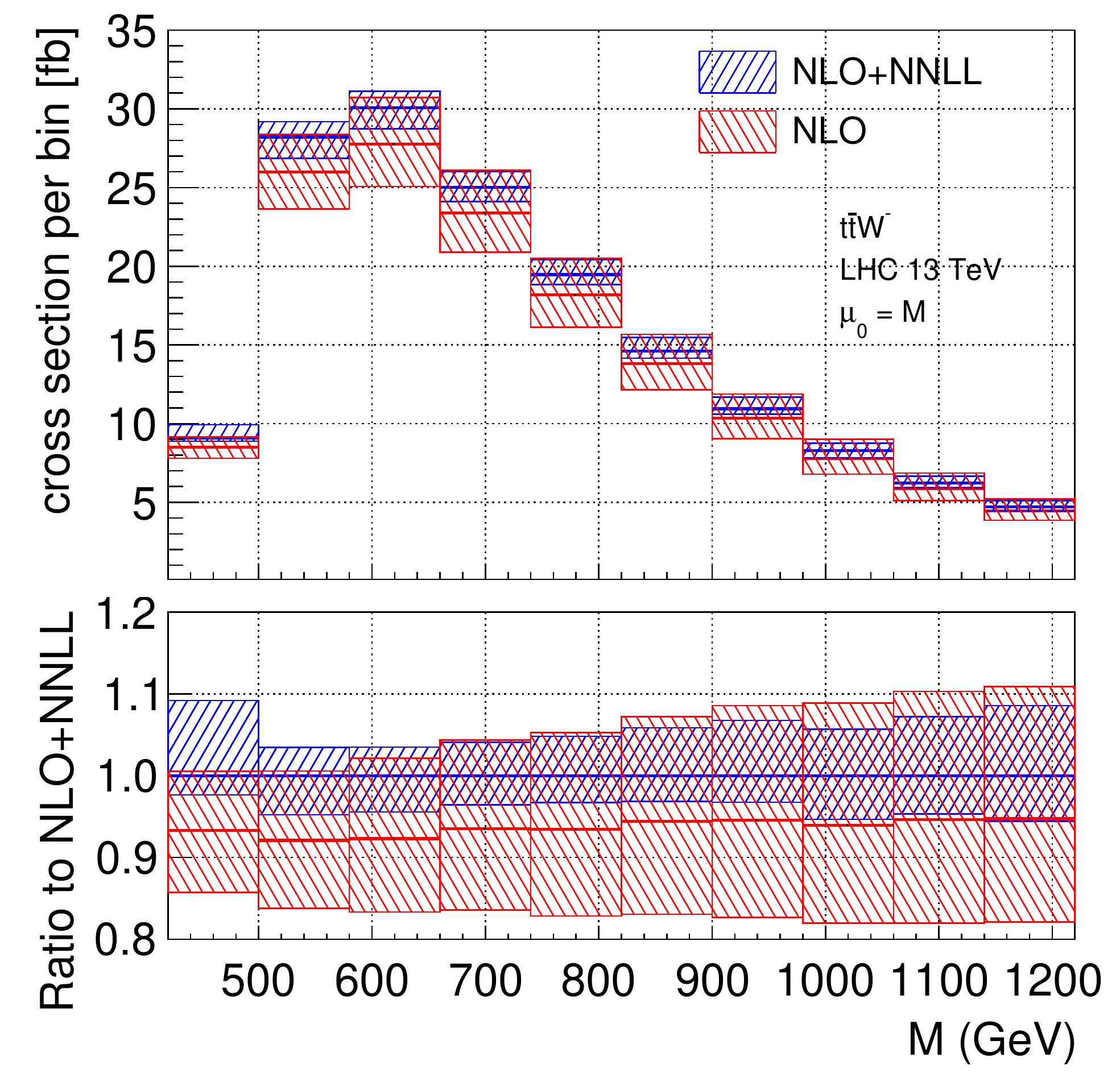} & \includegraphics[width=7.2cm]{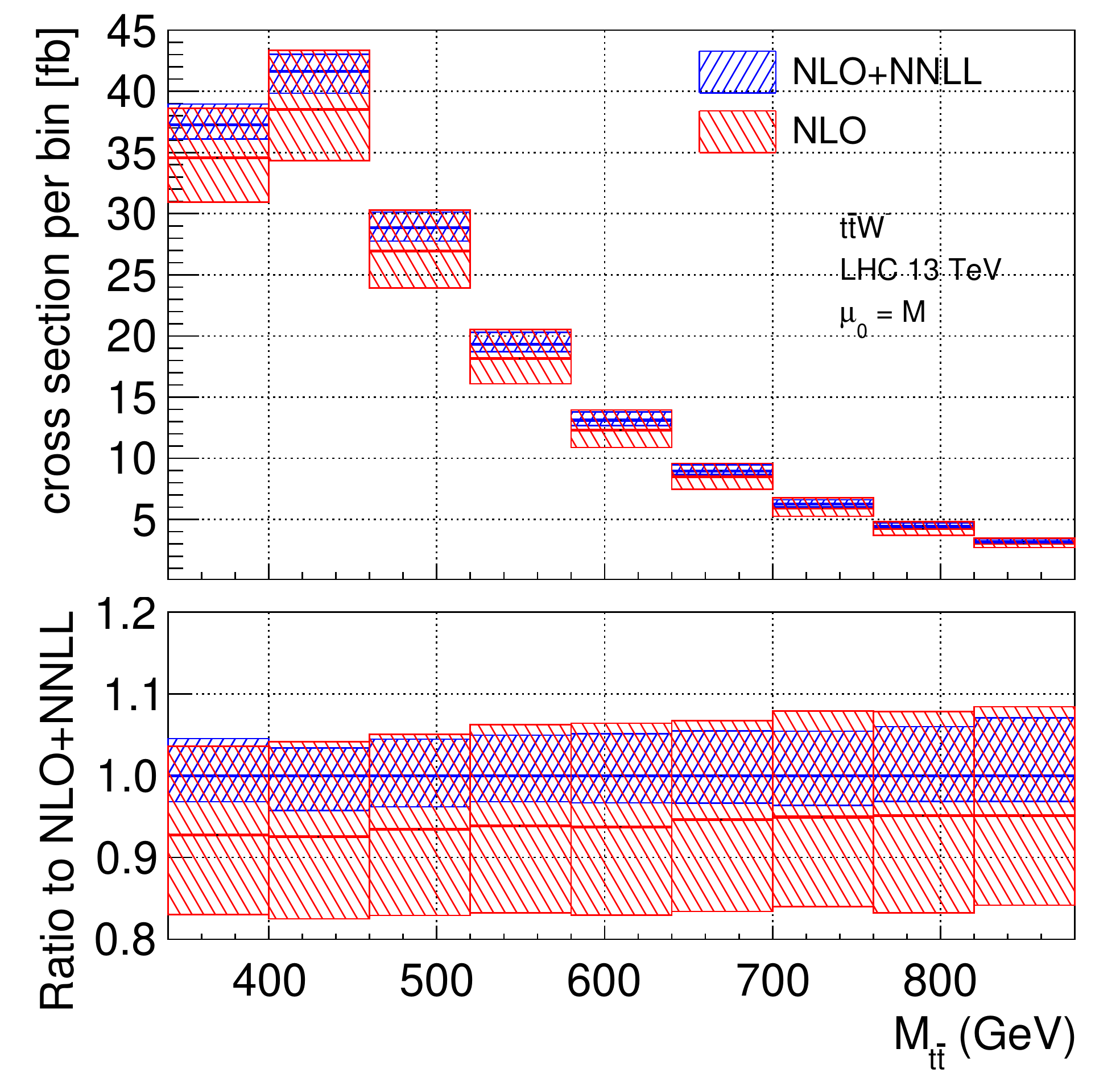} \\
			\includegraphics[width=7.2cm]{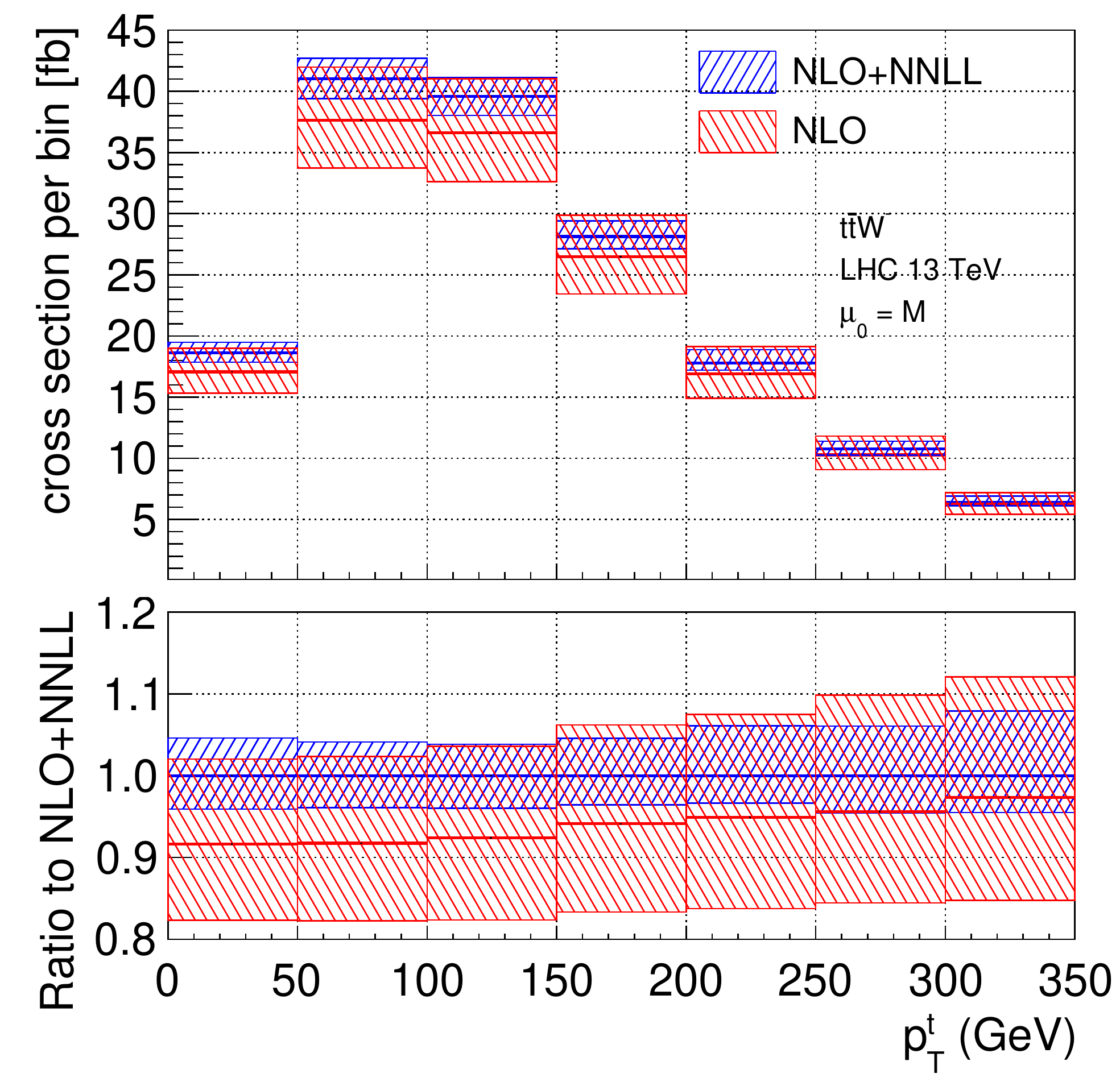} & \includegraphics[width=7.2cm]{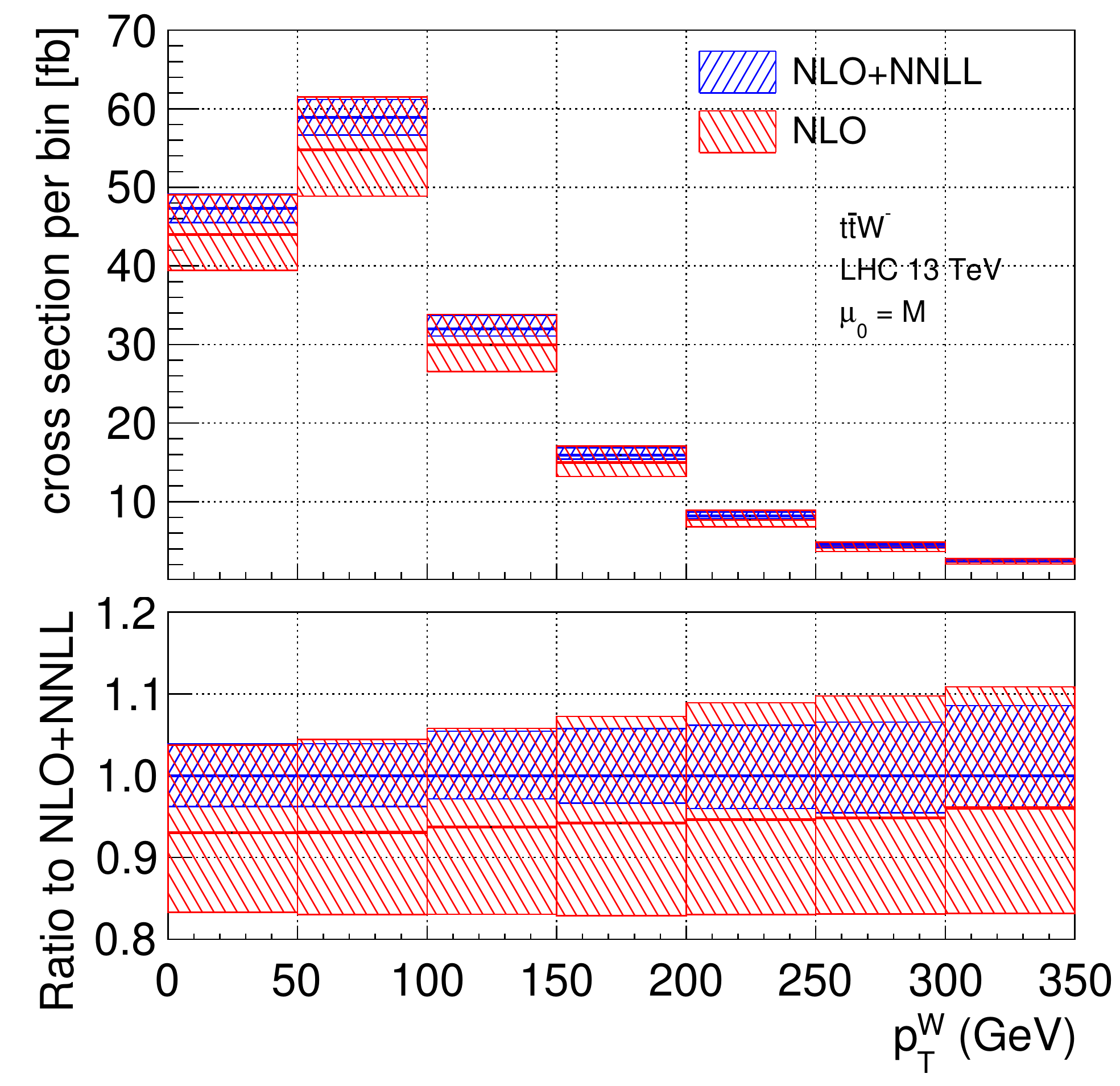} \\
		\end{tabular}
	\end{center}
	\caption{$t \bar{t} W^-$ production at $\sqrt{s} = 13$ TeV: Differential distributions at NLO+NNLL (blue bands) compared to the  NLO calculation (red bands). MMHT 2014 NNLO PDFs were used for the NLO+NNLL calculation, while the NLO calculation was carried out with  MMHT 2014 NLO PDFs.
		\label{fig:Wm13NNLL}
	}
\end{figure}

The same features discussed above for the case of $t \bar{t} W^+$ production at $\sqrt{s} = 8$~TeV are observed in the case of $t \bar{t} W^-$ production at $\sqrt{s} = 8$~TeV and for the same two processes for the LHC operating at a center of mass energy of $13$~TeV. Our best predictions for the differential distributions considered in this work are the ones obtained through NLO+NNLL calculations. For this reason we conclude this section by repeating the  analysis of Figure~\ref{fig:Wp8NLOvsNNLL} also for the case of  $t \bar{t} W^-$ production at $\sqrt{s} = 8$~TeV (Figure~\ref{fig:Wm8NNLL}), $t \bar{t} W^+$ at $\sqrt{s} = 13$~TeV (Figure~\ref{fig:Wp13NNLL}), and $t \bar{t} W^-$ at $\sqrt{s} = 13$~TeV (Figure~\ref{fig:Wm13NNLL}).

\section{Conclusions}
\label{sec:conclusions}

In this paper we studied the resummation of the soft gluon emission corrections to the associated production of a top-quark pair and a $W$ boson at the LHC. After analyzing the factorization of the partonic cross section in the partonic threshold limit in Mellin space, we collected all of the elements needed in order to implement the resummation of these corrections to NNLL accuracy.
The numerical evaluation of the NNLL resummation formula was carried out by means of an in-house parton level Monte Carlo program. This program can be employed to evaluate arbitrary differential distributions depending on the momenta of the massive particles in the final state. In order to validate our method and to test the Monte-Carlo code, we evaluated the total cross section and four different differential distributions. While in this paper we did not include cuts on the final state phase space, arbitrary cuts on the momenta of the final state particles can be introduced in a straightforward way. Additional work along the lines of \cite{Broggio:2014yca} would allow one to generalize the program in order to account for the decay of the massive particles.
 
In this way we obtained predictions for the $t \bar{t} W^\pm$ production total cross section and differential distributions which are valid to NNLL accuracy and are matched to NLO calculations carried out with \mgamc. These NLO+NNLL predictions are the main result of this paper. Since a full evaluation of the complete NNLO corrections to this process is for the moment out of reach, the calculations presented here  represent the most precise predictions for $t \bar{t} W^\pm$ production available at the moment of writing.
This aspect is particularly relevant given the fact that this process has already been measured at the LHC both in Run I and in Run II.
The effect of the NLO+NNLL corrections is to increase the total cross section and differential distributions to the upper part of the uncertainty intervals identified by NLO calculations through scale variation. The residual perturbative uncertainty affecting NLO+NNLL predictions  is, as expected, smaller than the NLO one in all cases, and in particular in the tails of the distributions.

The calculations and analyses carried out in this work also serve as testing ground for the study of processes such as $t \bar{t} H$ and $t \bar{t} Z$, which are crucial to the LHC
physics program. While these processes share many features with the associated production of a top pair and a $W$ boson, they are more complicated because they involve not one but two partonic channels in the soft limit; namely the quark-annihilation channel (also present in $t \bar{t} W^\pm$) and the 
gluon-fusion channel
(absent in $t \bar{t} W^\pm$). For these reasons the evaluation of $t \bar{t} H$ and 
$t \bar{t} Z$ is computationally more expensive from the point of view of running time; consequently, it made sense to develop and optimize our method and the in-house parton level Monte-Carlo code by studying $t \bar{t} W^\pm$ production. We plan to turn to the calculation of  $t \bar{t} H$ and $t \bar{t} Z$ production to NLO+NNLL accuracy in future work.

\section*{Acknowledgments}
We would like to thank P.~Maierhofer and S.~Pozzorini for their assistance in the use of {\tt OpenLoops}.
The in-house Monte Carlo code which we developed and employed to
evaluate the (differential) cross sections presented in this paper was
run on the computer cluster of the Center for Theoretical Physics at
the Physics Department of New York City College of Technology.\\
A.~B. and B.~P. would like to thank the Physics Department of the New York City College of Technology and A.~F. and B.~P. the Physics Department T31 of the Technische Universit\"at M\"unchen for their kind hospitality at different stages of this project.
The work of A.~F. and G.~O. is supported in part by the National Science Foundation under Grant No. PHY-1417354. 

\bibliography{mybib}

\providecommand{\href}[2]{#2}\begingroup\raggedright\begin{thebibliography}{10}

\bibitem{Khachatryan:2015sha}
{\scshape CMS} collaboration, V.~Khachatryan et~al., \emph{{Observation of top
  quark pairs produced in association with a vector boson in pp collisions at $
  \sqrt{s}=8 $ TeV}},
  \href{http://dx.doi.org/10.1007/JHEP01(2016)096}{\emph{JHEP} {\bf 01} (2016)
  096}, [\href{http://arxiv.org/abs/1510.01131}{{\tt 1510.01131}}].

\bibitem{Aad:2015eua}
{\scshape ATLAS} collaboration, G.~Aad et~al., \emph{{Measurement of the $
  t\overline{t}W $ and $ t\overline{t}Z $ production cross sections in pp
  collisions at $ \sqrt{s}=8 $ TeV with the ATLAS detector}},
  \href{http://dx.doi.org/10.1007/JHEP11(2015)172}{\emph{JHEP} {\bf 11} (2015)
  172}, [\href{http://arxiv.org/abs/1509.05276}{{\tt 1509.05276}}].

\bibitem{Garzelli:2012bn}
M.~V. Garzelli, A.~Kardos, C.~G. Papadopoulos and Z.~Trocsanyi,
  \emph{{t$\bar{t}$$W^{\pm}$ and t$\bar{t}$Z Hadroproduction at NLO accuracy in
  QCD with Parton Shower and Hadronization effects}},
  \href{http://dx.doi.org/10.1007/JHEP11(2012)056}{\emph{JHEP} {\bf 11} (2012)
  056}, [\href{http://arxiv.org/abs/1208.2665}{{\tt 1208.2665}}].

\bibitem{Campbell:2012dh}
J.~M. Campbell and R.~K. Ellis, \emph{{$t\bar{t}W^{\pm}$ production and decay
  at NLO}}, \href{http://dx.doi.org/10.1007/JHEP07(2012)052}{\emph{JHEP} {\bf
  07} (2012) 052}, [\href{http://arxiv.org/abs/1204.5678}{{\tt 1204.5678}}].

\bibitem{Maltoni:2014zpa}
F.~Maltoni, M.~L. Mangano, I.~Tsinikos and M.~Zaro, \emph{{Top-quark charge
  asymmetry and polarization in $t\overline{t}W^±$ production at the LHC}},
  \href{http://dx.doi.org/10.1016/j.physletb.2014.07.033}{\emph{Phys. Lett.}
  {\bf B736} (2014) 252--260}, [\href{http://arxiv.org/abs/1406.3262}{{\tt
  1406.3262}}].

\bibitem{ATLAS:13TeV}
{\scshape ATLAS} collaboration, G.~Aad et~al., \emph{{Measurement of the
  $t\bar{t}Z$ and $t\bar{t}W$ production cross sections in multilepton final
  states using 3.2 fb$^{-1}$ of $pp$ collisions at 13 TeV at the LHC}},
  {\emph{ATLAS-CONF-2016-003} (2016) }.

\bibitem{Becher:2014oda}
T.~Becher, A.~Broggio and A.~Ferroglia, \emph{{Introduction to Soft-Collinear
  Effective Theory}}, vol.~896.
\newblock Springer, 2015,
  \href{http://dx.doi.org/10.1007/978-3-319-14848-9}{10.1007/978-3-319-14848-9}.

\bibitem{Broggio:2015lya}
A.~Broggio, A.~Ferroglia, B.~D. Pecjak, A.~Signer and L.~L. Yang,
  \emph{{Associated production of a top pair and a Higgs boson beyond NLO}},
  \href{http://dx.doi.org/10.1007/JHEP03(2016)124}{\emph{JHEP} {\bf 03} (2016)
  124}, [\href{http://arxiv.org/abs/1510.01914}{{\tt 1510.01914}}].

\bibitem{Cascioli:2011va}
F.~Cascioli, P.~Maierhofer and S.~Pozzorini, \emph{{Scattering Amplitudes with
  Open Loops}},
  \href{http://dx.doi.org/10.1103/PhysRevLett.108.111601}{\emph{Phys. Rev.
  Lett.} {\bf 108} (2012) 111601}, [\href{http://arxiv.org/abs/1111.5206}{{\tt
  1111.5206}}].

\bibitem{Denner:2002ii}
A.~Denner and S.~Dittmaier, \emph{{Reduction of one loop tensor five point
  integrals}},
  \href{http://dx.doi.org/10.1016/S0550-3213(03)00184-6}{\emph{Nucl. Phys.}
  {\bf B658} (2003) 175--202}, [\href{http://arxiv.org/abs/hep-ph/0212259}{{\tt
  hep-ph/0212259}}].

\bibitem{Denner:2005nn}
A.~Denner and S.~Dittmaier, \emph{{Reduction schemes for one-loop tensor
  integrals}},
  \href{http://dx.doi.org/10.1016/j.nuclphysb.2005.11.007}{\emph{Nucl. Phys.}
  {\bf B734} (2006) 62--115}, [\href{http://arxiv.org/abs/hep-ph/0509141}{{\tt
  hep-ph/0509141}}].

\bibitem{Denner:2010tr}
A.~Denner and S.~Dittmaier, \emph{{Scalar one-loop 4-point integrals}},
  \href{http://dx.doi.org/10.1016/j.nuclphysb.2010.11.002}{\emph{Nucl. Phys.}
  {\bf B844} (2011) 199--242}, [\href{http://arxiv.org/abs/1005.2076}{{\tt
  1005.2076}}].

\bibitem{Denner:2014gla}
A.~Denner, S.~Dittmaier and L.~Hofer, \emph{{COLLIER - A fortran-library for
  one-loop integrals}}, {\emph{PoS} {\bf LL2014} (2014) 071},
  [\href{http://arxiv.org/abs/1407.0087}{{\tt 1407.0087}}].

\bibitem{Denner:2016kdg}
A.~Denner, S.~Dittmaier and L.~Hofer, \emph{{Collier: a fortran-based Complex
  One-Loop LIbrary in Extended Regularizations}},
  \href{http://arxiv.org/abs/1604.06792}{{\tt 1604.06792}}.

\bibitem{Ossola:2007ax}
G.~Ossola, C.~G. Papadopoulos and R.~Pittau, \emph{{CutTools: A Program
  implementing the OPP reduction method to compute one-loop amplitudes}},
  \href{http://dx.doi.org/10.1088/1126-6708/2008/03/042}{\emph{JHEP} {\bf 03}
  (2008) 042}, [\href{http://arxiv.org/abs/0711.3596}{{\tt 0711.3596}}].

\bibitem{Cullen:2011ac}
G.~Cullen, N.~Greiner, G.~Heinrich, G.~Luisoni, P.~Mastrolia, G.~Ossola et~al.,
  \emph{{Automated One-Loop Calculations with GoSam}},
  \href{http://dx.doi.org/10.1140/epjc/s10052-012-1889-1}{\emph{Eur. Phys. J.}
  {\bf C72} (2012) 1889}, [\href{http://arxiv.org/abs/1111.2034}{{\tt
  1111.2034}}].

\bibitem{Cullen:2014yla}
G.~Cullen et~al., \emph{{G$\scriptsize{O}$S$\scriptsize{AM}$-2.0: a tool for
  automated one-loop calculations within the Standard Model and beyond}},
  \href{http://dx.doi.org/10.1140/epjc/s10052-014-3001-5}{\emph{Eur. Phys. J.}
  {\bf C74} (2014) 3001}, [\href{http://arxiv.org/abs/1404.7096}{{\tt
  1404.7096}}].

\bibitem{Binoth:2008uq}
T.~Binoth, J.~P. Guillet, G.~Heinrich, E.~Pilon and T.~Reiter, \emph{{Golem95:
  A Numerical program to calculate one-loop tensor integrals with up to six
  external legs}},
  \href{http://dx.doi.org/10.1016/j.cpc.2009.06.024}{\emph{Comput. Phys.
  Commun.} {\bf 180} (2009) 2317--2330},
  [\href{http://arxiv.org/abs/0810.0992}{{\tt 0810.0992}}].

\bibitem{Mastrolia:2010nb}
P.~Mastrolia, G.~Ossola, T.~Reiter and F.~Tramontano, \emph{{Scattering
  AMplitudes from Unitarity-based Reduction Algorithm at the Integrand-level}},
  \href{http://dx.doi.org/10.1007/JHEP08(2010)080}{\emph{JHEP} {\bf 08} (2010)
  080}, [\href{http://arxiv.org/abs/1006.0710}{{\tt 1006.0710}}].

\bibitem{Peraro:2014cba}
T.~Peraro, \emph{{Ninja: Automated Integrand Reduction via Laurent Expansion
  for One-Loop Amplitudes}},
  \href{http://dx.doi.org/10.1016/j.cpc.2014.06.017}{\emph{Comput. Phys.
  Commun.} {\bf 185} (2014) 2771--2797},
  [\href{http://arxiv.org/abs/1403.1229}{{\tt 1403.1229}}].

\bibitem{Hirschi:2011pa}
V.~Hirschi, R.~Frederix, S.~Frixione, M.~V. Garzelli, F.~Maltoni and R.~Pittau,
  \emph{{Automation of one-loop QCD corrections}},
  \href{http://dx.doi.org/10.1007/JHEP05(2011)044}{\emph{JHEP} {\bf 05} (2011)
  044}, [\href{http://arxiv.org/abs/1103.0621}{{\tt 1103.0621}}].

\bibitem{Hirschi:2016mdz}
V.~Hirschi and T.~Peraro, \emph{{Tensor integrand reduction via Laurent
  expansion}}, \href{http://dx.doi.org/10.1007/JHEP06(2016)060}{\emph{JHEP}
  {\bf 06} (2016) 060}, [\href{http://arxiv.org/abs/1604.01363}{{\tt
  1604.01363}}].

\bibitem{Pecjak:2016nee}
B.~D. Pecjak, D.~J. Scott, X.~Wang and L.~L. Yang, \emph{{Resummed differential
  cross sections for top-quark pairs at the LHC}},
  \href{http://dx.doi.org/10.1103/PhysRevLett.116.202001}{\emph{Phys. Rev.
  Lett.} {\bf 116} (2016) 202001}, [\href{http://arxiv.org/abs/1601.07020}{{\tt
  1601.07020}}].

\bibitem{Ferroglia:2015ivv}
A.~Ferroglia, B.~D. Pecjak, D.~J. Scott and L.~L. Yang, \emph{{QCD resummations
  for boosted top production}}, {\emph{PoS} {\bf TOP2015} (2016) 052},
  [\href{http://arxiv.org/abs/1512.02535}{{\tt 1512.02535}}].

\bibitem{Li:2014ula}
H.~T. Li, C.~S. Li and S.~A. Li, \emph{{Renormalization group improved
  predictions for $t\bar{t}W^\pm$ production at hadron colliders}},
  \href{http://dx.doi.org/10.1103/PhysRevD.90.094009}{\emph{Phys. Rev.} {\bf
  D90} (2014) 094009}, [\href{http://arxiv.org/abs/1409.1460}{{\tt
  1409.1460}}].

\bibitem{Alwall:2014hca}
J.~Alwall, R.~Frederix, S.~Frixione, V.~Hirschi, F.~Maltoni, O.~Mattelaer
  et~al., \emph{{The automated computation of tree-level and next-to-leading
  order differential cross sections, and their matching to parton shower
  simulations}}, \href{http://dx.doi.org/10.1007/JHEP07(2014)079}{\emph{JHEP}
  {\bf 07} (2014) 079}, [\href{http://arxiv.org/abs/1405.0301}{{\tt
  1405.0301}}].

\bibitem{Ferroglia:2009ep}
A.~Ferroglia, M.~Neubert, B.~D. Pecjak and L.~L. Yang, \emph{{Two-loop
  divergences of scattering amplitudes with massive partons}},
  \href{http://dx.doi.org/10.1103/PhysRevLett.103.201601}{\emph{Phys. Rev.
  Lett.} {\bf 103} (2009) 201601}, [\href{http://arxiv.org/abs/0907.4791}{{\tt
  0907.4791}}].

\bibitem{Ferroglia:2009ii}
A.~Ferroglia, M.~Neubert, B.~D. Pecjak and L.~L. Yang, \emph{{Two-loop
  divergences of massive scattering amplitudes in non-abelian gauge theories}},
  \href{http://dx.doi.org/10.1088/1126-6708/2009/11/062}{\emph{JHEP} {\bf 11}
  (2009) 062}, [\href{http://arxiv.org/abs/0908.3676}{{\tt 0908.3676}}].

\bibitem{Ahrens:2010zv}
V.~Ahrens, A.~Ferroglia, M.~Neubert, B.~D. Pecjak and L.~L. Yang,
  \emph{{Renormalization-Group Improved Predictions for Top-Quark Pair
  Production at Hadron Colliders}},
  \href{http://dx.doi.org/10.1007/JHEP09(2010)097}{\emph{JHEP} {\bf 09} (2010)
  097}, [\href{http://arxiv.org/abs/1003.5827}{{\tt 1003.5827}}].

\bibitem{Mastrolia:2012bu}
P.~Mastrolia, E.~Mirabella and T.~Peraro, \emph{{Integrand reduction of
  one-loop scattering amplitudes through Laurent series expansion}},
  \href{http://dx.doi.org/10.1007/JHEP11(2012)128,
  10.1007/JHEP06(2012)095}{\emph{JHEP} {\bf 06} (2012) 095},
  [\href{http://arxiv.org/abs/1203.0291}{{\tt 1203.0291}}].

\bibitem{vanDeurzen:2013saa}
H.~van Deurzen, G.~Luisoni, P.~Mastrolia, E.~Mirabella, G.~Ossola and
  T.~Peraro, \emph{{Multi-leg One-loop Massive Amplitudes from Integrand
  Reduction via Laurent Expansion}},
  \href{http://dx.doi.org/10.1007/JHEP03(2014)115}{\emph{JHEP} {\bf 03} (2014)
  115}, [\href{http://arxiv.org/abs/1312.6678}{{\tt 1312.6678}}].

\bibitem{Catani:1996yz}
S.~Catani, M.~L. Mangano, P.~Nason and L.~Trentadue, \emph{{The Resummation of
  soft gluons in hadronic collisions}},
  \href{http://dx.doi.org/10.1016/0550-3213(96)00399-9}{\emph{Nucl. Phys.} {\bf
  B478} (1996) 273--310}, [\href{http://arxiv.org/abs/hep-ph/9604351}{{\tt
  hep-ph/9604351}}].

\bibitem{Moch:2005ba}
S.~Moch, J.~A.~M. Vermaseren and A.~Vogt, \emph{{Higher-order corrections in
  threshold resummation}},
  \href{http://dx.doi.org/10.1016/j.nuclphysb.2005.08.005}{\emph{Nucl. Phys.}
  {\bf B726} (2005) 317--335}, [\href{http://arxiv.org/abs/hep-ph/0506288}{{\tt
  hep-ph/0506288}}].

\bibitem{Becher:2007ty}
T.~Becher, M.~Neubert and G.~Xu, \emph{{Dynamical Threshold Enhancement and
  Resummation in Drell-Yan Production}},
  \href{http://dx.doi.org/10.1088/1126-6708/2008/07/030}{\emph{JHEP} {\bf 07}
  (2008) 030}, [\href{http://arxiv.org/abs/0710.0680}{{\tt 0710.0680}}].

\bibitem{Bonvini:2012az}
M.~Bonvini, S.~Forte, M.~Ghezzi and G.~Ridolfi, \emph{{Threshold Resummation in
  SCET vs. Perturbative QCD: An Analytic Comparison}},
  \href{http://dx.doi.org/10.1016/j.nuclphysb.2012.04.010}{\emph{Nucl. Phys.}
  {\bf B861} (2012) 337--360}, [\href{http://arxiv.org/abs/1201.6364}{{\tt
  1201.6364}}].

\bibitem{Bonvini:2014qga}
M.~Bonvini, S.~Forte, G.~Ridolfi and L.~Rottoli, \emph{{Resummation
  prescriptions and ambiguities in SCET vs. direct QCD: Higgs production as a
  case study}}, \href{http://dx.doi.org/10.1007/JHEP01(2015)046}{\emph{JHEP}
  {\bf 01} (2015) 046}, [\href{http://arxiv.org/abs/1409.0864}{{\tt
  1409.0864}}].

\bibitem{Harland-Lang:2014zoa}
L.~A. Harland-Lang, A.~D. Martin, P.~Motylinski and R.~S. Thorne, \emph{{Parton
  distributions in the LHC era: MMHT 2014 PDFs}},
  \href{http://dx.doi.org/10.1140/epjc/s10052-015-3397-6}{\emph{Eur. Phys. J.}
  {\bf C75} (2015) 204}, [\href{http://arxiv.org/abs/1412.3989}{{\tt
  1412.3989}}].

\bibitem{Broggio:2014yca}
A.~Broggio, A.~S. Papanastasiou and A.~Signer, \emph{{Renormalization-group
  improved fully differential cross sections for top pair production}},
  \href{http://dx.doi.org/10.1007/JHEP10(2014)098}{\emph{JHEP} {\bf 10} (2014)
  98}, [\href{http://arxiv.org/abs/1407.2532}{{\tt 1407.2532}}].

\end{thebibliography}\endgroup
	
\bibliographystyle{JHEP}

\end{document}